\documentclass{jpp}
\usepackage{graphicx}

\usepackage[utf8]{inputenc}
\usepackage[T1]{fontenc}
\usepackage{amsmath,amsfonts,amssymb}

\usepackage{multirow}
\usepackage{tablefootnote}
\usepackage{nicefrac}

\shorttitle{Plasma Equilibrium in Diamagnetic Trap with NBI}
\shortauthor{M. S. Khristo and A. D. Beklemishev}

\title{Plasma Equilibrium in Diamagnetic Trap with Neutral Beam Injection}

\author{Mikhail S. Khristo\aff{1, 2}
  \corresp{\email{khristo.mikhail@gmail.com}},
  \and Alexei D. Beklemishev\aff{1, 2}}

\affiliation{\aff{1} Budker Institute of Nuclear Physics, 11, Acad. Lavrentieva Pr., Novosibirsk, 630090, Russia
\aff{2} Novosibirsk State University, 1, Pirogova str., Novosibirsk, 630090, Russia}

\begin{document}

\maketitle

\begin{abstract}
This paper presents a theoretical model of plasma equilibrium in the
diamagnetic confinement mode in an axisymmetric mirror device with
neutral beam injection. The hot ionic component is described within
the framework of the kinetic theory, since the Larmor radius of the
injected ions appears to be comparable to or even larger than the
characteristic scale of the magnetic field inhomogeneity. The electron
drag of the hot ions is taken into account, while the angular scattering
due to ion-ion collisions is neglected. The background warm plasma,
on the contrary, is considered to be in local thermal equilibrium,
i.e. has a Maxwellian distribution function and is described in terms
of magnetohydrodynamics. The density of the hot ions is assumed to
be negligible compared to that of the warm plasma. Both the conventional
gas-dynamic loss and the non-adiabatic loss specific to the diamagnetic
confinement mode are taken into account. In this work, we do not consider
the effects of the warm plasma rotation as well as the inhomogeneity
of the electrostatic potential. A self-consistent theoretical model
of the plasma equilibrium is constructed. In the case of the cylindrical
bubble, this model is reduced to a simpler one. The numerical solutions
in the limit of a thin transition layer of the diamagnetic bubble
are found. Examples of the equilibria corresponding to the GDMT device
are considered.
\end{abstract}

\section{Introduction}

\textit{Diamagnetic confinement}, or \textit{diamagnetic bubble} \citep{Beklemishev2016,Khristo2019,Khristo2022,Kotelnikov2020,Chernoshtanov2020,Chernoshtanov2022b,Soldatkina2023},
is a new operating mode designed to significantly enhance confinement
in linear systems \citep{Ivanov2017,Dimov2005,Steinhauer2011a}.
The main idea of this regime is to form a high-pressure plasma \textit{bubble}
in the central part of a mirror device. Inside such a bubble, the
plasma pressure reaches the equilibrium limit corresponding to $\beta\rightarrow1$,
and the magnetic field tends to zero due to the plasma diamagnetism.
The lifetime of the particles in the \textit{diamagnetic trap}, i.e.
the mirror device operating in the diamagnetic confinement mode, is
estimated by \citet{Beklemishev2016} in the framework of magnetohydrodynamics
(MHD) as follows:
\[
\tau_{\mathrm{DC}}\sim\sqrt{\tau_{\mathrm{GD}}\tau_{\perp}},
\]
where $\tau_{\mathrm{GD}}$ is the lifetime in the gas-dynamic trap
(GDT) \citep{Ivanov2017}, and $\tau_{\perp}$ is the transverse
transport time. Since plasma transport across the magnetic field in
axisymmetric traps is usually far below axial losses: $\tau_{\perp}\gg\tau_{\mathrm{GD}}$,
a significantly enhanced overall confinement is expected in the diamagnetic
trap: $\tau_{\mathrm{DC}}\sim\tau_{\mathrm{GD}}\sqrt{\tau_{\perp}/\tau_{\mathrm{GD}}}\gg\tau_{\mathrm{GD}}$.
In this regard, there is a practical interest in the experimental
and theoretical study of this mode. In particular, one of the goals
of the new generation linear machine Gas-Dynamic Multiple-Mirror Trap
(GDMT) \citep{Beklemishev2013,Bagryansky2019,Skovorodin2023} is
to experimentally verify the concept of the diamagnetic confinement. In
addition, the study of the diamagnetic regime is planned at the Compact
Axisymmetric Toroid (CAT) \citep{Bagryansky2016} currently
operating at the Budker Institute of Nuclear Physics (Budker INP).

One can find a number of earlier theoretical works related to $\beta\sim1$
plasma confinement in open traps \citep[see][]{Grad1967,Newcomb1981,Lansky1993,Lotov1996,Kotelnikov2010,Kotelnikov2011}.
Effective confinement of plasma with $\beta\sim1$ in a gas-dynamic
system was demonstrated experimentally in the 2MK-200 test facility
by \citet{Zhitlukhin1984}. High $\beta$ plasma confinement was also
studied in the so-called magnetoelectrostatic traps \citep[see][]{Pastukhov1978,Pastukhov1980,Ioffe1981,Pastukhov2021}.
Structures similar to the diamagnetic bubble, called magnetic holes,
are often observed in space plasmas \citep[see][]{Kaufmann1970,Turner1977,Tsurutani2011,Kuznetsov2015}. 

The idea of diamagnetic confinement was originally proposed by \citet{Beklemishev2016}.
The possibility of a discharge transition into the diamagnetic
confinement mode is briefly discussed, and a stationary MHD model
of a diamagnetic bubble equilibrium is constructed in the cylindrical
approximation. Further, this hydrodynamic equilibrium model was extended
by \citet{Khristo2019,Khristo2022} to the case of a non-paraxial axisymmetric
trap. In particular, diamagnetic bubble equilibria in the GDMT are
computed, and the effect of magnetic field corrugation on equilibrium
in a diamagnetic trap is studied. \citet{Beklemishev2016} and \citet{Khristo2022}
also briefly discuss the possibility of MHD stabilization by a combination
of the vortex confinement \citep[see][]{Soldatkina2008,Beklemishev2010,Bagryansky2011}
and the conducting wall \citep[see][]{Kaiser1985,Berk1987,Kotelnikov2022}.

As already noted, the magnetic field inside the diamagnetic bubble is
close to zero. In this case, the Larmor radius and the mean free path
of high-energy particles can be comparable to or even exceed the characteristic
scale of magnetic field inhomogeneity, which is beyond the scope of
MHD theory. Therefore, there is a need for a detailed kinetic model
of fast particles in the diamagnetic trap. Research on the kinetic theory
of the diamagnetic regime is already underway at the Budker INP. \citet{Kotelnikov2020}
constructed a fully kinetic equilibrium model in cylindrical geometry
with a distribution function isotropic in the transverse plane inside
the bubble. The collisionless dynamics of individual particles in
a diamagnetic trap were studied by \citet{Chernoshtanov2022b}. In
addition, a particle-in-cell model for simulating high-pressure plasma
in an open trap is currently being developed in the Budker INP by
\citet{Kurshakov2023}. Work is also underway to create a code for
numerical simulation of the diamagnetic regime \citep[see][]{Boronina2020,Efimova2020,Chernoshtanov2023,Chernoshtanov2024}.

Modern experiments on linear devices, such as the already mentioned
GDMT and CAT, commonly involve the injection of high-energy (on the
order of several tens of electronvolts) neutral beams to heat the
plasma \citep{Ivanov2017,Belchenko2018}. With this in mind, in the
present paper, we aim to construct a theoretical model of diamagnetic
bubble equilibrium in a gas-dynamic trap with neutral beam injection.
The neutral beams are absorbed by the background plasma at a temperature
much lower than the energy of the injected atoms. For this reason,
to describe the equilibrium in such a system, we assume the plasma
to consist of two fractions: the background \textit{warm plasma} and
the \textit{hot ions} resulting from the neutral beam injection. The former
is considered to be in local thermodynamic equilibrium and is described
in terms of MHD; we take as a basis the hydrodynamic model constructed
in the earlier works \citep{Beklemishev2016,Khristo2019,Khristo2022}.
The hot ions, on the contrary, are assumed to have a non-Maxwellian
distribution function and are described within the framework of the
kinetic theory. Similar hybrid equilibria were considered earlier
in application to field reversed configurations (FRCs) \citep{Rostoker2002,Qerushi2002a,Qerushi2002,Qerushi2003,Steinhauer2011b}. 

For simplicity, we consider the approximation of a long, axisymmetric
diamagnetic trap. In this case, in the absence of dissipation and
scattering, the azimuthal angular momentum and the total energy of
a single charged particle are conserved. At the same time, since the
magnetic field in the diamagnetic confinement regime has considerable
gradients, the magnetic moment $\mu=mv_{\perp}^{2}/2B$ is not globally
conserved. However, it was shown independently by \citet{Chernoshtanov2020,Chernoshtanov2022b}
and \citet{Kotelnikov2020} that in a diamagnetic trap, the adiabatic
invariant $I_{r}=\left(2\pi\right)^{-1}\oint p_{r}dr$ can be conserved
under specific conditions. In addition, it is known that in such axisymmetric
systems there is a region in phase space where particles are \textit{absolutely
confined} \citep[see][]{Morozov1966,Lovelace1978,Larrabee1979,Hsiao1985}.

In the present paper, we assume the hot ions to be injected into the
region of the phase space, where, first, the absolute confinement
criterion is met and, second, the condition of the adiabaticity is
violated. The former saves us from solving the complex problem of
taking into account the hot ion loss. The latter leads to the distribution
function of the hot ions being homogeneous on the hypersurface of
the constant azimuthal angular momentum and total energy, and hence
not depending on the adiabatic invariant $I_{r}$. Violation of the
adiabaticity occurs typically due to the non-paraxial ends of the
bubble configuration \citep{Beklemishev2016}. We also assume the
energy of the hot ions to be much higher than the temperature of the
warm plasma. In this approximation, the hot ions are mainly slowing
down on the electrons of the warm plasma and hardly collide with the
ions. On this basis, we completely neglect the angular scattering
when calculating the hot ion distribution function and take into account
only the weak drag force from the warm electrons.

The article is structured as follows: Section \ref{sec:basic_assumptions_of_theoretical_model}
focuses on the detailed problem statement; Section \ref{sec:magnetic_field}
defines the equations describing the equilibrium of the magnetic field;
Section \ref{sec:warm_plasma_equilibrium} derives the equilibrium
equations for the warm plasma. Section \ref{sec:hot_ion_equilibrium}
finds the equilibrium distribution function of the hot ions; Section
\ref{sec:cylindrical_bubble_model} reduces the theoretical model
to the case of the cylindrical diamagnetic bubble; Section \ref{sec:thin_transition_layer_limit}
is devoted to the solution of the equilibrium equations in the approximation
of a thin transition layer at the bubble boundary; Section \ref{sec:diamagnetic_confinement_in_gdmt}
considers examples of equilibria corresponding to the diamagnetic
confinement regime in the GDMT device; Section \ref{sec:summary}
summarizes the main results of this paper and discusses the issues
to be addressed in future work.

\section{Basic assumptions of theoretical model\label{sec:basic_assumptions_of_theoretical_model}}

Consider the stationary equilibrium of a diamagnetic bubble in an
axisymmetric gas-dynamic trap. At the periphery of the bubble, the
magnetic field is close to the \textit{vacuum magnetic field} $B_{\mathrm{v}}$,
i.e. the magnetic field without plasma. In the interior of the bubble,
the magnetic field is vanishingly small $B\ll B_{\mathrm{v}}$, being
almost completely expelled by diamagnetic plasma; this region we
further refer to as the \textit{core of the diamagnetic bubble}. We
denote the radius of the bubble core as $r_{0}$, which, generally
speaking, can vary along the trap: $r_{0}=r_{0}\left(z\right)$\footnote{Henceforth, cylindrical coordinates $\left(r,\theta,z\right)$ are
used; the corresponding unit vectors are denoted by a 'hat': $\hat{\boldsymbol{r}}$,
$\hat{\boldsymbol{\theta}}$, $\hat{\boldsymbol{z}}$.}. The core radius in the central section
of the trap, $z=0$, we define as $a=r_{0}\left(0\right)$. The region
at the boundary of the bubble, inside which the magnetic field changes
from $B\simeq0$ in the core to $B=B_{\mathrm{v}}$ at the periphery,
we further refer to as the \textit{transition layer of the diamagnetic
bubble}. 

Let the plasma consist of \textit{hot ions}, resulting from the neutral
beam injection, and background \textit{warm plasma}. The warm plasma
is assumed to be in local thermal equilibrium with the temperature
$T=T\left(r,z\right)$ and described in terms of MHD. The hot ions,
on the contrary, are expected to have a non-Maxwellian distribution
function and are described within the framework of kinetic theory.

As was previously found by \citet{Khristo2022}, the specific choice
of the warm plasma transport model in the bubble core seems to have little effect
on the equilibrium. For this reason, we further assume that, inside
the bubble core, the magnetic field is identically zero $B\equiv0$,
and the warm plasma electrical conductivity $\sigma_{\mathrm{w}}$ and transverse
diffusion coefficient $\mathcal{D}$ are extremely high: $\sigma_{\mathrm{w}}\rightarrow\infty$
and $\mathcal{D}\rightarrow\infty$. In what follows, we also consider
the approximation of a long paraxial bubble with short non-paraxial
ends. In addition, we do not take into account the rotation of the
warm plasma and the effect of the electrostatic potential inhomogeneity.
The warm plasma radial electric currents are also neglected. We understand
that these issues are definitely important and thus should be addressed
in future work.

In an axisymmetric system, the azimuthal canonical angular momentum
$\mathcal{P}$ and the total energy $\mathcal{E}$ of a charged particle
are conserved:
\begin{equation}
\mathcal{P}=m_{\mathrm{s}}rv_{\theta}+\frac{e_{\mathrm{s}}}{c}\frac{\psi}{2\pi}=\mathrm{const},\quad\mathcal{E}=\frac{m_{\mathrm{s}}v^{2}}{2}=\mathrm{const},\label{eq:P_E}
\end{equation}
where $m_{\mathrm{s}}$, $e_{\mathrm{s}}$ are the mass and the electric
charge of a particle of species $\mathrm{s}$, respectively, $v_{\theta}$
is the azimuthal component of the particle velocity, $\psi=\int_{0}^{r}B_{z}\left(r',z\right)2\pi r'dr'$
is the flux of the longitudinal ('poloidal') component of the magnetic
field\footnote{In the axisymmetric case, one can fix the vector potential gauge:
$\partial_{\theta}A=0$. Then the azimuthal component of the vector
potential is $A_{\theta}=\psi/2\pi r$.}. Assuming the magnetic flux in the mirrors to be approximately equal
to $\psi_{\mathrm{m}}\simeq B_{\mathrm{m}}\pi r^{2}$, we arrive at
the \textit{absolute confinement} criterion in the form \citep[see][]{Morozov1966,Lovelace1978,Larrabee1979,Hsiao1985}:

\begin{equation}
-\mathcal{R}\Omega_{\mathrm{s}}\mathcal{P}>\mathcal{E},\label{eq:absolute_confinement}
\end{equation}
where $\Omega_{\mathrm{s}}=e_{\mathrm{s}}B_{\mathrm{v}}/m_{\mathrm{s}}c$
is the cyclotron frequency in the vacuum magnetic field of the central
section of the trap $B_{\mathrm{v}}$, $\mathcal{R}=B_{\mathrm{m}}/B_{\mathrm{v}}$
is the vacuum mirror ratio, and $B_{\mathrm{m}}$ is the mirror magnetic
field. Worth noting is that the particles are absolutely confined
only if the direction of their rotation and the direction of the Larmor
rotation are the same. In other words, positively charged particles
with $\Omega_{\mathrm{s}}>0$ are confined if $\mathcal{P}<0$, and
negatively charged particles with $\Omega_{\mathrm{s}}<0$ are confined
if $\mathcal{P}>0$. In order to avoid solving the complex problem
of taking into account the hot ion loss, in this paper, we consider
the neutral beam being injected into the absolute confinement region
(\ref{eq:absolute_confinement}). It is clear that, in practice, the
injection should be carried out in this way to avoid undesirable loss
of the hot ions.

It was discovered by \citet{Kotelnikov2020} and \citet{Chernoshtanov2020,Chernoshtanov2022b}
that there exists the adiabatic invariant $I_{r}=\left(2\pi\right)^{-1}\oint p_{r}dr$
for particles in a diamagnetic trap. As additionally shown by \citet{Chernoshtanov2020,Chernoshtanov2022b},
$I_{r}$ is conserved for particles with not too high velocity $v_{\parallel}$
along the magnetic field, namely, the adiabaticity criterion can be
approximately written as a limitation on the pitch angle $\xi$:
\begin{equation}
\tan\xi=\frac{v_{\perp}}{\left|v_{\parallel}\right|}\gtrsim\left|\frac{dr_{0}}{dz}\right|_{\max},\label{eq:adiabaticity_criterion}
\end{equation}
where $v_{\perp}$ is the velocity transverse to the magnetic field,
$\left|dr_{0}/dz\right|_{\max}$ is the maximum inclination of the
field line corresponding to the bubble core boundary $r_{0}=r_{0}\left(z\right)$.
The criterion (\ref{eq:adiabaticity_criterion}) imposes considerable
restrictions on the uniformity of the magnetic field. In addition
to the requirement of sufficient global smoothness of the longitudinal
profile of the field lines, the small-scale ripples of the magnetic
field, resulting from the discreteness of the magnetic system and
instabilities, should also be insignificant. Due to this, the region
in the phase space where the criterion (\ref{eq:adiabaticity_criterion})
is met usually proves to be relatively narrow. To simplify the theoretical
model, we further assume that the condition (\ref{eq:adiabaticity_criterion})
is violated and the adiabatic invariant $I_{r}$ is not conserved.
As a result, the particle dynamics becomes chaotic, and the trajectories
fill the hypersurfaces with constant angular momentum and energy $\mathcal{P}=\mathrm{const}$,
$\mathcal{E}=\mathrm{const}$. Based on this, we approximately consider
the distribution function of the hot ions depending only on $\mathcal{P}$
and $\mathcal{E}$. The collisionless dynamics of the hot ions in
a diamagnetic trap may resemble that in FRCs. In particular, \citet{Landsman2004}
provide a classification of particle trajectories in FRCs, where the
regions of phase space corresponding to chaotic motion are observed
as well.

As shown by \citet{Chernoshtanov2020,Chernoshtanov2022b}, in a diamagnetic
trap there should also arise an additional axial loss, the so-called
\textit{non-adiabatic loss}. It results from the particles escaping
the trap through a 'leak' in the phase space outside the region of
the absolute confinement (\ref{eq:absolute_confinement}). The characteristic
non-adiabatic loss time for plasma with a Maxwellian distribution
function at the center of the trap can be estimated as
\[
\tau_{\mathrm{s}\parallel0}\sim\tau_{\mathrm{GD\,s}}\frac{a}{\rho_{\mathrm{s}}},
\]
where $\tau_{\mathrm{GD\,s}}=\mathcal{R}L/v_{T\mathrm{s}}$ is the
gas-dynamic confinement time, $a$ and $L$ are the characteristic
radial size and length of the diamagnetic bubble, $\rho_{\mathrm{s}}=v_{T\mathrm{s}}/\left|\Omega_{\mathrm{s}}\right|$
is the characteristic Larmor radius, and $v_{T\mathrm{s}}=\sqrt{T/m_{\mathrm{s}}}$
is the thermal velocity. Note that the time $\tau_{\mathrm{s}\parallel0}$
does not depend on the particle mass since it is canceled in $\left|\Omega_{\mathrm{s}}\right|/v_{T\mathrm{s}}^{2}$.
As can be seen, despite the additional non-adiabatic loss, confinement
of the warm plasma in the diamagnetic trap can still be significantly
enhanced compared to the 'classical' gas-dynamic trap, provided the
ratio $\rho_{\mathrm{s}}/a$ is sufficiently small. 

In this work, we assume the injection energy $\mathcal{E}_{\mathrm{NB}}$
to be large compared to the temperature of the warm plasma $T$. However,
due to the significant difference in masses, the characteristic velocity
of the injected ions $\sqrt{\mathcal{E}_{\mathrm{NB}}/m_{\mathrm{i}}}$
is still much less than the thermal velocity of the warm electrons
$\sqrt{T/m_{\mathrm{e}}}$. Summarizing the above, we have:
\[
1\ll\frac{\mathcal{E}_{\mathrm{NB}}}{T}\ll\frac{m_{\mathrm{i}}}{m_{\mathrm{e}}},
\]
where $m_{\mathrm{e}}$ and $m_{\mathrm{i}}$ are the electron and
ion masses, respectively. We also consider the density\footnote{For brevity, in this article, the \textit{number density}, i.e. the
number of particles per unit volume, is referred to as simply \textit{density}.} of the hot ions to be negligible compared to that of the warm plasma,
so the quasi-neutrality condition is satisfied for the latter: $n_{\mathrm{i}}\simeq n_{\mathrm{e}}/Z$,
where $n_{\mathrm{e}}$ and $n_{\mathrm{i}}$ are the densities of
warm electrons and ions, respectively, and $Z$ is the ion atomic
number. In addition, we neglect collisions of the hot ions with ions
and take into account only the drag force from the warm electrons.
In other words, we use a simple linear kinetic equation for the hot
ions, in which we keep only the terms related to the slowing down
of the hot ions on the warm electrons, neglecting the collisional angular
scattering. This approximation apparently corresponds to the high-energy
limit $\mathcal{E}\gg T$ of the injected ions. 

Let us briefly summarize the formulation of the problem. We consider
the stationary axisymmetric equilibrium of a diamagnetic trap in the
paraxial approximation. The plasma consists of the warm plasma in thermal
equilibrium and non-Maxwellian hot ions. The warm plasma is described
in terms of MHD. Both gas-dynamic and non-adiabatic losses of the
warm plasma are taken into account. The electrical conductivity and transverse
diffusion coefficient of the warm plasma inside the bubble core are
assumed to be extremely high. We neglect the rotation of the warm
plasma and the inhomogeneity of the electrostatic potential. Hot ions
are considered within the framework of kinetic theory. A linear kinetic
equation is used, which takes into account only the electron drag
of the hot ions. The distribution function of the hot ions is considered
to depend only on azimuthal angular momentum and energy.

\section{Magnetic field equilibrium distribution\label{sec:magnetic_field}}

The equilibrium stationary distribution of the magnetic field can
be found from Maxwell's equation with a source:
\begin{equation}
\nabla\times\boldsymbol{B}=4\pi c^{-1}\boldsymbol{J},\label{eq:maxwell}
\end{equation}
where $\boldsymbol{J}$ is the total electric current density. Since
the radial and longitudinal currents, as well as the azimuthal magnetic
field, are assumed to be zero, we have $\boldsymbol{J}=J_{\theta}\hat{\boldsymbol{\theta}}$
and $\boldsymbol{A}=A_{\theta}\hat{\boldsymbol{\theta}}$. In the
axisymmetric case before us, we can also fix the vector potential
gauge as follows:
\[
\partial_{\theta}A_{\theta}=0\quad\Leftrightarrow\quad A_{\theta}=A_{\theta}\left(r,z\right).
\]
Then the vector potential $A_{\theta}$ proves to be related to the
axial magnetic flux $\psi$:
\[
A_{\theta}=\frac{\psi}{2\pi r},\quad\psi=\int\limits _{0}^{r}B_{z}\left(r',z\right)2\pi r'dr',
\]
Therefore, the equation (\ref{eq:maxwell}) reduces to the following
two-dimensional second-order differential equation for $\psi$:
\begin{equation}
r\partial_{r}\left(r^{-1}\partial_{r}\psi\right)+\partial_{z}^{2}\psi=-8\pi^{2}c^{-1}rJ_{\theta},\label{eq:grad_shafranov}
\end{equation}
which is the Grad-Shafranov equilibrium equation \citep{Grad1958,Shafranov1958}
in the case of an axisymmetric mirror device. The electric current
density $J_{\theta}$ on the right-hand side of the equation (\ref{eq:grad_shafranov})
contains both the plasma diamagnetic current $J_{\mathrm{p}\theta}$
and the current in the external coils $J_{\mathrm{v}\theta}$, i.e.
$J_{\theta}=J_{\mathrm{p}\theta}+J_{\mathrm{v}\theta}$. Having solved
the equation (\ref{eq:grad_shafranov}), one can evaluate the magnetic
field $\boldsymbol{B}$ via the found magnetic flux $\psi$ as follows:
\[
\boldsymbol{B}=\frac{1}{2\pi r}\nabla\psi\times\hat{\boldsymbol{\theta}},
\]
and then
\[
B=\left|\boldsymbol{B}\right|=\sqrt{\left(\partial_{r}\psi\right)^{2}+\left(\partial_{z}\psi\right)^{2}}.
\]

We consider the external boundary condition of the equation (\ref{eq:grad_shafranov})
to be the regularity of the magnetic field at infinity:
\begin{equation}
\left.B\right|_{r^{2}+z^{2}\rightarrow+\infty}\rightarrow B_{\infty},\quad0\leq B_{\infty}<+\infty.\label{eq:field_external}
\end{equation}
This corresponds to a plasma with a free boundary, i.e. confined only
due to the magnetic field of the external coils. At the same time,
by definition, there is no magnetic field in the interior of the bubble
core, i.e. $B\equiv0$, and hence $\psi\equiv0$, for $r\leq r_{0}\left(z\right)$,
where $r_{0}=r_{0}\left(z\right)$ is the bubble core radius. Then
one can set the following internal boundary conditions:
\begin{gather}
\left.B\right|_{r=r_{0}}=0,\label{eq:field_internal}\\
\left.\psi\right|_{r=r_{0}}=0.\label{eq:core_radius}
\end{gather}
Relations (\ref{eq:field_external}) and (\ref{eq:field_internal})
together define the boundary condition for the equation (\ref{eq:grad_shafranov}).
In turn, relation (\ref{eq:core_radius}) determines the bubble core
boundary $r_{0}=r_{0}\left(z\right)$.

The total plasma electric current density $J_{\mathrm{p}\theta}$
is the sum of the warm plasma current density $J_{\mathrm{w}\theta}$
and the current density of the hot ions $J_{\mathrm{h}\theta}$. The
former is to be found by solving hydrodynamic equilibrium equations
for the warm plasma; as a basis, we take the MHD models constructed
in previous works \citep[see][]{Beklemishev2016,Khristo2019,Khristo2022}.
The latter is determined by the distribution function of the hot ions,
which should be found from the solution of the corresponding kinetic
equation. These two issues are dealt with in the following Sections:
\ref{sec:warm_plasma_equilibrium} and \ref{sec:hot_ion_equilibrium},
respectively.

\section{Warm plasma equilibrium\label{sec:warm_plasma_equilibrium}}

As mentioned above, we consider the warm plasma to be in local thermal
equilibrium. We also assume the hot ion density to be negligible compared
to the density of the warm plasma. This allows us to consider the
warm plasma as quasi-neutral: $n_{\mathrm{i}}\simeq n_{\mathrm{e}}/Z$,
where $n_{\mathrm{e}}$ and $n_{\mathrm{i}}$ are the densities of
the warm electrons and ions, respectively, and $Z$ is the ion atomic
number. In addition, we completely neglect the warm plasma rotation
and the electrostatic potential inhomogeneity. The radial and longitudinal
electric currents, as well as the azimuthal magnetic field, are set
equal to zero.

\subsection{Force equilibrium}

The electric current density of the warm plasma $\boldsymbol{J}_{\mathrm{w}}$,
which appears in equation (\ref{eq:grad_shafranov}), can be found
from the force balance equation for the warm plasma\footnote{The inertial forces acting on the warm plasma, such as centrifugal
force, are neglected in this equation. The proper description of the
azimuthal force balance is related to the angular momentum equilibrium
and, consequently, to the warm plasma rotation, the accounting of
which is beyond the scope of the present article. }:
\begin{equation}
\nabla p=c^{-1}\boldsymbol{J}_{\mathrm{w}}\times\boldsymbol{B},\label{eq:force_balance}
\end{equation}
where $p$ is the warm plasma pressure\footnote{In what follows, the warm plasma pressure $p$ is assumed to be isotropic,
which corresponds to the case of a filled loss cone.}. It can be seen that, due to the stationary axisymmetric equilibrium
being considered, the sum of the azimuthal friction forces acting
on the entire warm plasma is equal to zero. 

It immediately follows from the equation (\ref{eq:force_balance})
that the warm plasma pressure $p$ is the constant on the flux surfaces
$\psi={\rm const}$ and, therefore, is a function of the magnetic
flux $\psi$ only: $p=p\left(\psi\right)$. In addition, due to the
high longitudinal electron thermal conductivity, the temperature of
the warm plasma $T$ appears to be constant along the field lines.
Together with axial symmetry, this results in the temperature also
being a flux function: $T=T\left(\psi\right)$. Eventually, assuming
the pressure, temperature, and density\footnote{In what follows, the \textit{density of the warm plasma} should be
understood as the \textit{density of the warm ions}.} of the warm plasma to be related by the equation of state:
\begin{equation}
p=\left(1+Z\right)n_{\mathrm{i}}T,\label{eq:state}
\end{equation}
we arrive at the same for the warm plasma density: $n_{\mathrm{i}}=n_{\mathrm{i}}\left(\psi\right)$. 

As a result, the equation (\ref{eq:force_balance}) allows the warm
plasma current density to be expressed in terms of the magnetic field
and pressure gradient:
\begin{equation}
J_{\mathrm{w}\theta}=2\pi rc\frac{dp}{d\psi}.\label{eq:warm_plasma_current}
\end{equation}
The magnetic flux distribution $\psi=\psi\left(r,z\right)$ is determined
by Maxwell's equations (\ref{eq:grad_shafranov}), and the pressure
profile $p=p\left(\psi\right)$ should be found from the solution
of the material and thermal equilibrium equations of the warm plasma.

\subsection{Material equilibrium}

To obtain the warm plasma equilibrium equation, we consider the material
balance in a flux tube $\psi=\mathrm{const}$:
\begin{equation}
2\Phi_{\mathrm{i}\parallel\mathrm{m}}+\Phi_{\mathrm{i}\perp}-W_{\mathrm{i}}=0,\label{eq:flow_balance}
\end{equation}
where $\Phi_{\mathrm{i}\parallel\mathrm{m}}$ is the axial loss of
the warm ions from the flux tube, determined by the longitudinal flow
through a mirror throat, $\Phi_{\mathrm{i}\perp}$ is the warm ion
flow transverse to the flux surface $\psi=\mathrm{const}$, and $W_{\mathrm{i}}=W_{\mathrm{i}}\left(\psi\right)$
is the total warm ion source inside the flux tube. In other words,
$W_{\mathrm{i}}$ represents the number of warm ion-electron pairs supplied by an
external source to the interior of the surface $\psi=\mathrm{const}$
per unit time. Longitudinal and transverse ion fluxes are determined
as follows:
\[
\Phi_{\mathrm{i}\parallel\mathrm{m}}=\int\limits _{S_{\mathrm{m}}\left(\psi\right)}n_{\mathrm{i}}\boldsymbol{u}_{\mathrm{i}}\cdot d\boldsymbol{S},\quad\Phi_{\mathrm{i}\perp}=\int\limits _{S_{\perp}\left(\psi\right)}n_{\mathrm{i}}\boldsymbol{u}_{\mathrm{i}}\cdot d\boldsymbol{S},
\]
where $\boldsymbol{u}_{\mathrm{i}}$ is the warm ion flow velocity.
Integration is carried out over the cross section of the magnetic
surface in the mirror $S_{\mathrm{m}}\left(\psi\right)$ and the flux
surface $S_{\perp}\left(\psi\right)$ between the mirrors, respectively. 

Gas dynamic confinement implies the axial mirror loss being described
by an outflow through the nozzle:
\begin{equation}
\Phi_{\mathrm{i}\parallel\mathrm{m}}=\int\limits _{S_{\mathrm{m}}\left(\psi\right)}n_{\mathrm{i}}u_{\mathrm{m}}dS,\label{eq:longitudinal_flow}
\end{equation}
where $u_{\mathrm{m}}\sim\sqrt{T/m_{\mathrm{i}}}$ is the warm plasma
flow velocity in the mirror throats. The ion flow velocity transverse
to the flux surface $\psi=\mathrm{const}$ can be found from the generalized
Ohm's law, which can be written in the form:
\begin{equation}
\frac{1}{c}\left[\boldsymbol{u}_{\mathrm{i}}\times\boldsymbol{B}\right]_{\theta}=\frac{J_{\mathrm{w}\theta}}{\sigma_{\mathrm{w}}},\label{eq:ohm_law}
\end{equation}
where $\sigma_{\mathrm{w}}$ is the electrical conductivity of the warm plasma\footnote{Generally speaking, the warm plasma electrical conductivity $\sigma_{\mathrm{w}}$
may differ from the 'classical' Spitzer conductivity $\sigma_{\mathrm{Sp}}=Z^{2}e^{2}n_{\mathrm{i}}/m_{\mathrm{i}}\nu_{\mathrm{ie}}$.
Exotic conditions of the diamagnetic confinement mode, such as, for
instance, sharp gradients of the warm plasma parameters and magnetic
field, may lead to anomalous transport.}. The force balance (\ref{eq:force_balance}) together with the azimuthal
projection of the Ohm's law (\ref{eq:ohm_law}) yields the warm plasma
transverse flow velocity:
\begin{equation}
\boldsymbol{u}_{\mathrm{i}\perp}=-\frac{c^{2}}{\sigma_{\mathrm{w}}B^{2}}\nabla p.\label{eq:u_i_perp}
\end{equation}
Then the transverse ion flux is
\begin{equation}
\Phi_{\mathrm{i}\perp}=-\Lambda_{\mathrm{i}\perp}\frac{dp}{d\psi},\quad\Lambda_{\mathrm{i}\perp}=4\pi^{2}c^{2}n_{\mathrm{i}}\int\limits _{\gamma_{\perp}\left(\psi\right)}\frac{r^{2}dl}{\sigma_{\mathrm{w}}B},\label{eq:transverse_flow}
\end{equation}
where integration is carried out in $\left(r,z\right)$ space along
the curve $\gamma_{\perp}\left(\psi\right)$ corresponding to the
flux-surface $\psi=\mathrm{const}$. 

Finally, substituting (\ref{eq:longitudinal_flow}) and (\ref{eq:transverse_flow})
into the material balance (\ref{eq:flow_balance}) and differentiating
it with respect to $\psi$, we obtain the warm ion material equilibrium
equation in the following form:
\begin{equation}
-\frac{d}{d\psi}\left(\Lambda_{\mathrm{i}\perp}\frac{dp}{d\psi}\right)+\frac{2u_{\mathrm{m}}}{B_{\mathrm{m}}}n_{\mathrm{i}}=\frac{dW_{\mathrm{i}}}{d\psi},\label{eq:flow_equilibrium}
\end{equation}
where $B_{\mathrm{m}}$ is the magnetic field in the mirror throats;
here we also take into account that $d\psi=B_{\mathrm{m}}dS_{\mathrm{m}}$. 

\subsection{Thermal equilibrium}

For a given distribution of the magnetic flux, the equilibrium state
of the warm plasma is fully described by three parameters: density
$n_{\mathrm{i}}$, temperature $T$ and pressure $p$. A closed system
of equations can be obtained by supplementing (\ref{eq:state}) and
(\ref{eq:flow_equilibrium}) with another equation relating these
three parameters. In the previous MHD models \citep[see][]{Beklemishev2016,Khristo2019,Khristo2022},
the plasma temperature $T$ is simply assumed to be constant. However,
introducing an equation describing the thermal equilibrium of the
warm plasma would be more proper. 

The law of thermodynamics for a moving warm plasma element reads:
\begin{equation}
\nabla\cdot\left(\frac{3}{2}p\boldsymbol{u}_{\mathrm{i}}\right)=\nabla\cdot\left(\varkappa_{\mathrm{w}}\nabla T\right)+\frac{J_{\mathrm{w}}^{2}}{\sigma_{\mathrm{w}}}+q_{\mathrm{h}}-p\nabla\cdot\boldsymbol{u}_{\mathrm{i}},\label{eq:thermodynamics_law}
\end{equation}
where $-\varkappa_{\mathrm{w}}\nabla T$ is the heat flux density
due to thermal conductivity with a coefficient $\varkappa_{\mathrm{w}}$,
$q_{\mathrm{h}}$ is the density of the heating power from the hot
ions (see expression (\ref{eq:q_h}) in Section \ref{sec:hot_ion_equilibrium}).
The left-hand side of the equation is the change in the internal energy
of the moving element with time. The first three terms on the right-hand
side describe the total heat release, and the last term is related
to the mechanical work. Here, we consider the heating to be due to
the neutral beam injection. If necessary, heating from other sources
can also be taken into account by simply adding the corresponding
terms to the right-hand side of the equation.

The equations (\ref{eq:force_balance}) and (\ref{eq:u_i_perp}) together
yield the resistive heating power density:
\[
\frac{J_{\mathrm{w}}^{2}}{\sigma_{\mathrm{w}}}=-\boldsymbol{u}_{\mathrm{i}}\cdot\nabla p.
\]
Substituting it into (\ref{eq:thermodynamics_law}) we arrive at
\[
\frac{5}{2}\nabla\cdot\left(p\boldsymbol{u}_{\mathrm{i}}\right)-\nabla\cdot\left(\varkappa_{\mathrm{w}}\nabla T\right)=q_{\mathrm{h}}.
\]
Integrating this equation over the volume of the flux tube $\psi=\mathrm{const}$
between the mirrors results in
\begin{equation}
2\Phi_{\mathrm{E}\parallel}+\Phi_{\mathrm{E}\perp}=Q_{\mathrm{h}},\label{eq:energy_flow_balance}
\end{equation}
where $\Phi_{\mathrm{E}\parallel}$ is the axial energy loss from
the flux tube,
\[
\Phi_{\mathrm{E}\perp}=\int\limits _{S_{\perp}\left(\psi\right)}\frac{5}{2}p\boldsymbol{u}_{\mathrm{i}}\cdot d\boldsymbol{S}-\int\limits _{S_{\perp}\left(\psi\right)}\varkappa_{\mathrm{w}}\nabla T\cdot d\boldsymbol{S}
\]
is the energy flow transverse to the flux surface $\psi=\mathrm{const}$,
and $Q_{\mathrm{h}}$ is the total heating power from the hot ions inside
the flux tube.

The axial energy loss in a gas-dynamic trap can be described as follows:
\[
\Phi_{\mathrm{E}\parallel}=\alpha_{\mathrm{E}}\int\limits _{S_{\mathrm{m}}\left(\psi\right)}Tn_{\mathrm{i}}u_{\mathrm{m}}dS.
\]
In other words, an ion-electron pair escaping the trap carries away
energy $\alpha_{\mathrm{E}}T$ through the mirror \citep[see][]{Ryutov2005,Skovorodin2019}.
The factor $\alpha_{\mathrm{E}}$ depends on the nature of the plasma
outflow through the mirror; for the GDT facility in Budker INP, this
parameter lies in the range of $6\div8$ \citep[see][]{Soldatkina2020}.
Further, taking into account (\ref{eq:u_i_perp}), the transverse
energy flow is written in the form:
\[
\Phi_{\mathrm{E}\perp}=\frac{5}{2}\frac{p}{n_{\mathrm{i}}}\Phi_{\mathrm{i}\perp}-\Lambda_{\mathrm{E}\perp}\frac{dT}{d\psi},
\]
where
\[
\Lambda_{\mathrm{E}\perp}=4\pi^{2}\int\limits _{\gamma_{\perp}\left(\psi\right)}\varkappa_{\mathrm{w}}Br^{2}dl.
\]

As a result, after taking the derivative of the energy balance (\ref{eq:energy_flow_balance})
with respect to $\psi$, we arrive at the thermal equilibrium equation
for the warm plasma in the following form:
\begin{equation}
-\frac{d}{d\psi}\left(\frac{5}{2}\frac{p}{n_{\mathrm{i}}}\Lambda_{\mathrm{i}\perp}\frac{dp}{d\psi}\right)-\frac{d}{d\psi}\left(\Lambda_{\mathrm{E}\perp}\frac{dT}{d\psi}\right)+\frac{2\alpha_{\mathrm{E}}u_{\mathrm{m}}}{B_{\mathrm{m}}}n_{\mathrm{i}}T=\frac{dQ_{\mathrm{h}}}{d\psi}.\label{eq:thermal_equilibrium}
\end{equation}

\subsection{Bubble core equilibrium}

Strictly speaking, equations (\ref{eq:warm_plasma_current}), (\ref{eq:flow_equilibrium}),
and (\ref{eq:thermal_equilibrium}) cannot be used to describe the
equilibrium of the warm plasma in the bubble core, where $\psi\rightarrow0$.
In addition, the magnetized plasma approximation obviously ceases
to be applicable there. For this reason, the equilibrium of the warm
plasma inside the bubble core should be considered separately.

Since the magnetic field inside the bubble core is close to zero,
the warm plasma transport in the core should increase significantly
compared to the external transverse diffusion. 
Therefore, we assume the pressure, the temperature, and the density
of the warm plasma to be uniform inside the entire bubble core. In addition, we further suppose the warm plasma in the bubble
core to be perfectly conducting, which results in the magnetic field
there being constant and identically equal to zero:
\[
B\equiv0,\quad r<r_{0}\left(z\right).
\]
This is possible only if there is no plasma current inside the core,
i.e. the electric current of the hot ions is completely compensated
by the inductive current of the warm plasma:
\begin{equation}
J_{\mathrm{w}\theta}=-J_{\mathrm{h}\theta},\quad r<r_{0}\left(z\right).\nonumber
\end{equation}

It is shown by \citet{Chernoshtanov2020,Chernoshtanov2022b} that the
so-called \textit{non-adiabatic loss} should arise in the diamagnetic
trap. In the case of a sufficiently collisional warm plasma with an
isotropic distribution function, there are particles outside the absolute
confinement region (\ref{eq:absolute_confinement}). These particles
may reach the mirrors and escape the trap directly from the bubble
core. The thermal equilibrium distribution of the warm ions in a diamagnetic
trap can be approximately represented as follows \citep[see][]{Chernoshtanov2020,Chernoshtanov2022b}:
\begin{equation}
f_{\mathrm{i}}\left(\mathcal{E},\mathcal{P}\right)=n_{\mathrm{i}0}\left(\frac{m_{\mathrm{i}}}{2\pi T_{0}}\right)^{\nicefrac{3}{2}}e^{-\frac{\mathcal{E}}{T_{0}}}\Theta\left(a-\frac{\left|\mathcal{P}\right|}{\sqrt{2m_{\mathrm{i}}\mathcal{E}}}\right),\label{eq:f_i}
\end{equation}
where $n_{\mathrm{i}0}$ and $T_{0}$ are the warm plasma density
and temperature in the bubble core, $\Theta\left(x\right)$ is the
Heaviside step function, $a=\max\left[r_{0}\left(z\right)\right]=r_{0}\left(0\right)$
is the maximum radius of the core, $\mathcal{P}$ and $\mathcal{E}$
are the canonical angular momentum and the total energy, respectively,
defined as in (\ref{eq:P_E}). Such a distribution function corresponds
to the collisional plasma when the characteristic Maxwellization time
is much less than the confinement time. It can be obtained by direct
averaging of the 'ordinary' Maxwellian distribution over the hypersurface
of the constant $\mathcal{P}$ and $\mathcal{E}$ (see the definition
of averaging (\ref{eq:averaging}) in Section \ref{sec:hot_ion_equilibrium}).
The total axial loss from the core of the bubble is determined by
the flow through the mirrors:
\begin{equation}
\Phi_{\mathrm{i}\parallel0}\simeq\frac{3}{4}n_{\mathrm{i}0}v_{T\mathrm{i}0}\frac{2\pi a\rho_{\mathrm{i}0}}{\mathcal{R}},\label{eq:Phi_i_parallel_0}
\end{equation}
where $v_{T\mathrm{i}0}=\sqrt{T_{0}/m_{\mathrm{i}}}$ is the warm
ion thermal velocity, $\rho_{\mathrm{i}0}=v_{T\mathrm{i}0}m_{\mathrm{i}}c/eZB_{\mathrm{v}}$
is the characteristic warm ion Larmor radius in the vacuum field $B_{\mathrm{v}}$,
$\mathcal{R}=B_{\mathrm{m}}/B_{\mathrm{v}}$ is the vacuum mirror
ratio, and $B_{\mathrm{m}}\simeq\mathrm{const}$ is the mirror magnetic
field. The expression (\ref{eq:Phi_i_parallel_0}) is obtained in
the limit $a\gg\rho_{\mathrm{i}0}/\mathcal{R}$; for details, refer
to Appendix \ref{sec:non_adiabatic_loss}.

The non-adiabatic loss (\ref{eq:Phi_i_parallel_0}) is to be taken
into account when setting the internal boundary condition for the
material equilibrium equation (\ref{eq:flow_equilibrium}). Namely,
the total transverse ion flow through the boundary of the bubble core
is the total warm ion source inside the core $W_{\mathrm{i}0}$ minus
the non-adiabatic loss through the mirrors $2\Phi_{\mathrm{i}\parallel0}$:
\begin{equation}
\left.\left(-\Lambda_{\mathrm{i}\perp}\frac{dp}{d\psi}\right)\right|_{\psi\rightarrow0}=W_{\mathrm{i}0}-2\Phi_{\mathrm{i}\parallel0}.\label{eq:flow_equilibrium_internal}
\end{equation}

The corresponding axial energy loss is defined as follows:
\[
\Phi_{\mathrm{E}\parallel0}=\alpha_{\mathrm{E}0}T_{0}\Phi_{\mathrm{i}\parallel0},
\]
where $\alpha_{\mathrm{E}0}$ is a coefficient similar to the previously
introduced $\alpha_{\mathrm{E}}$ for gas-dynamic energy loss. Then
the internal boundary condition for the thermal equilibrium equation
(\ref{eq:thermal_equilibrium}) is
\begin{equation}
\left.\left(-\frac{5}{2}\frac{p}{n_{\mathrm{i}}}\Lambda_{\mathrm{i}\perp}\frac{dp}{d\psi}-\Lambda_{\mathrm{E}\perp}\frac{dT}{d\psi}\right)\right|_{\psi\rightarrow0}=Q_{\mathrm{h}0}-2\alpha_{\mathrm{E}0}T_{0}\Phi_{\mathrm{i}\parallel0},\label{eq:thermal_equilibrium_internal}
\end{equation}
where $Q_{\mathrm{h}0}$ is the total heat release from the hot ions inside
the bubble core.

To complete the formulation of the boundary value problem, the external
boundary conditions are to be set. As such, for example, the conditions
\begin{equation}
\left.n_{\mathrm{i}}\right|_{r=a_{\mathrm{lim}}}=0,\quad\left.T\right|_{r=a_{\mathrm{lim}}}=0\label{eq:equilibrium_external}
\end{equation}
can be chosen, which correspond to an external limiter located at
the radius $a_{\mathrm{lim}}$. If the particle source is localized
inside the bubble core, as we assume in the present paper, due to
the large axial loss, the actual boundary of the warm plasma typically
does not reach the limiter. This means that in this case, the equilibrium
appears to be weakly dependent on the external boundary conditions
(\ref{eq:equilibrium_external}).

\section{Hot ion equilibrium\label{sec:hot_ion_equilibrium}}

Consider the dynamics of a single ion with charge $Ze$ and mass $m_{{\rm i}}$
in a given axisymmetric magnetic field, which is determined by the
magnetic flux distribution $\psi=\psi\left(r,z\right)$ (see Section
\ref{sec:magnetic_field}). Then the equations of motion have the
form:
\begin{align*}
\dot{r} & =\frac{P_{r}}{m_{\mathrm{i}}}, & \dot{P}_{r} & =\frac{P_{\theta}-\Psi}{m_{\mathrm{i}}r^{2}}\left(\frac{P_{\theta}-\Psi}{r}+\partial_{r}\Psi\right),\\
\dot{\theta} & =\frac{P_{\theta}-\Psi}{m_{\mathrm{i}}r^{2}}, & \dot{P}_{\theta} & =0,\\
\dot{z} & =\frac{P_{z}}{m_{\mathrm{i}}}, & \dot{P}_{z} & =\frac{P_{\theta}-\Psi}{m_{\mathrm{i}}r^{2}}\partial_{z}\Psi,
\end{align*}
where $\left(r,\theta,z,P_{r},P_{\theta},P_{z}\right)$ is a set of
canonically conjugate Hamiltonian variables, overdot indicates a time
derivative, and for brevity, the dimensionless magnetic flux $\Psi=Ze\psi/2\pi c$
is introduced. 

There are two global regular integrals of motion: the total energy
\[
\mathcal{E}=\frac{P_{r}^{2}}{2m_{\mathrm{i}}}+\frac{\left(P_{\theta}-\Psi\right)^{2}}{2m_{\mathrm{i}}r^{2}}+\frac{P_{z}^{2}}{2m_{\mathrm{i}}}
\]
and the angular momentum
\[
\mathcal{P}=P_{\theta}=m_{\mathrm{i}}r^{2}\dot{\theta}+\Psi.
\]
Excluding the cyclic variable $\theta$, for a given $\mathcal{P}$,
the number of degrees of freedom is reduced to two, and the ion dynamics
is equivalent to the motion of a particle with energy $\mathcal{E}$
in two-dimensional $\left(r,z\right)$ space in the effective potential
\begin{equation}
\varphi_{\mathrm{eff}}=\frac{\left(\mathcal{P}-\Psi\right)^{2}}{2m_{\mathrm{i}}r^{2}}.\label{eq:phi_eff}
\end{equation}
Therefore, the conservation of $\mathcal{P}$ and $\mathcal{E}$ leads
to the ion moving in a bounded area: $\mathcal{E}-\varphi_{\mathrm{eff}}>0.$ 

For given $\mathcal{E}$ and $\mathcal{P}$ the equality $\mathcal{E}=\varphi_{\mathrm{eff}}$
determines a curve in $\left(r,z\right)$ space corresponding to the
boundary of the region of ion motion. Then, taking into account that the
magnetic flux in the mirrors is approximately $\psi_{\mathrm{m}}=B_{\mathrm{m}}\pi r^{2}$,
we can find the radial boundaries in the mirror throats from the equation
\[
\left(\frac{\mathcal{R}\Omega_{\mathrm{i}}}{2}\right)^{2}\left(m_{\mathrm{i}}r^{2}\right)^{2}-2\left(\frac{\mathcal{R}\Omega_{\mathrm{i}}}{2}\mathcal{P}+\mathcal{E}\right)\left(m_{\mathrm{i}}r^{2}\right)+\mathcal{P}^{2}=0,
\]
where $\Omega_{\mathrm{i}}=ZeB_{\mathrm{v}}/m_{\mathrm{i}}c$ is the
ion cyclotron frequency in the vacuum magnetic field $B_{\mathrm{v}}$.
If this equation has no real roots, then the region of ion motion
does not reach the mirrors, and the ion is confined absolutely. This
brings us to the absolute confinement criterion (\ref{eq:absolute_confinement})
for the ions: 
\[
\left(\frac{\mathcal{R}\Omega_{\mathrm{i}}}{2}\mathcal{P}+\mathcal{E}\right)^{2}-\left(\frac{\mathcal{R}\Omega_{\mathrm{i}}}{2}\right)^{2}\mathcal{P}^{2}<0,\quad\Rightarrow\quad-\mathcal{R}\Omega_{\mathrm{i}}\mathcal{P}>\mathcal{E}.
\]
Otherwise, an ion outside the absolute confinement region may escape
the trap in a finite time.

\begin{figure}
\begin{centering}
\includegraphics{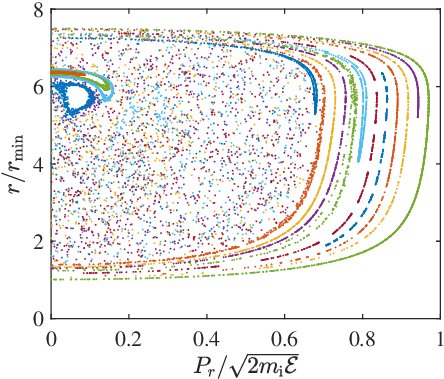}
\par\end{centering}
\caption{An example of the ion Poincaré map in the central plane, $z=0$, for
various initial conditions and fixed $\mathcal{E}$ and $\mathcal{P}$.
The magnetic field distribution is taken from MHD simulations for the GDMT configuration \citep[see][]{Khristo2022}. Vacuum magnetic field: $B_{\mathrm{v}}\simeq1.5\,\mathrm{T}$;
ion Larmor radius in the vacuum field: $\rho_{\mathcal{E}}=\sqrt{2m_{\mathrm{i}}\mathcal{E}}c/ZeB_{\mathrm{v}}\simeq1.6\,\mathrm{cm}$;
minimum distance from the trap axis that the ion can approach: $r_{\min}=\left|\mathcal{P}\right|/\sqrt{2m_{\mathrm{i}}\mathcal{E}}\simeq3.5\,\mathrm{cm}$.
\label{fig:poincare_map}}
\end{figure}

If there is a third conserved quantity in addition to $\mathcal{E}$
and $\mathcal{P}$, the system is integrable, and the dynamics of
the ion is regular. In particular, it was found by \citet{Chernoshtanov2020,Chernoshtanov2022b}
that for the ion with not too high longitudinal velocity, satisfying
the condition (\ref{eq:adiabaticity_criterion}), there is an adiabatic
invariant $I_{r}=\oint p_{r}dr$. Therefore, the trajectory of such
an ion is fully described by the given $\mathcal{E}$, $\mathcal{P}$,
and $I_{r}$. Otherwise, if the criterion (\ref{eq:adiabaticity_criterion})
is violated, there is no adiabatic invariant $I_{r}$, and the dynamics
of the ion becomes chaotic. This means that the ion trajectory ergodicly
fills a finite volume of phase space on the invariant hypersurface
of constant $\mathcal{E}$ and $\mathcal{P}$ \citep{Sagdeev1988,Lichtenberg1992}. 

The chaotic behavior of ions can be qualitatively explained by collisionless
scattering on the longitudinal inhomogeneities of the magnetic field
on the bubble boundary. 'Reflecting' from the magnetic field at the
boundary of the bubble leads to a change in the pitch angle of an
ion by approximately $\Delta\xi\sim\arctan\left|dr_{0}/dz\right|$.
Therefore, the adiabaticity criterion (\ref{eq:adiabaticity_criterion})
is met when the maximum scattering angle is not too large: $\left(\Delta\xi\right)_{\max}/\xi\lesssim1$,
while large-angle scattering: $\left(\Delta\xi\right)_{\max}/\xi\gtrsim1$,
on the considerable field inhomogeneities: $\left|dr_{0}/dz\right|\sim1$,
may result in violation of adiabaticity and hence to dynamic chaos.
With this in mind, we can also suppose that the characteristic time
of ergodization appears to be on the order of several free pass times
between the large-scale inhomogeneities: $\left|dr_{0}/dz\right|\sim1$.

In the presence of non-paraxial regions, where $\left|dr_{0}/dz\right|\sim1$,
which seems to be typical for the diamagnetic bubble equilibrium \citep{Kotelnikov2010,Kotelnikov2011,Beklemishev2016,Khristo2019,Khristo2022},
the adiabaticity criterion (\ref{eq:adiabaticity_criterion}) seems
to be violated for the majority of the injected ions. In particular,
Figure \ref{fig:poincare_map} illustrates an example of the ion Poincaré
map in the central plane, $z=0$, for various initial conditions and
fixed energy $\mathcal{E}$ and angular momentum $\mathcal{P}$. The
magnetic field distribution is taken from the MHD simulations of the diamagnetic bubble equilibrium in GDMT \citep[see][]{Khristo2022}. Vacuum magnetic field
in the central plane is $B_{\mathrm{v}}\simeq1.5\,\mathrm{T}$, the
ion Larmor radius in the vacuum magnetic field is $\rho_{\mathcal{E}}=\sqrt{2m_{\mathrm{i}}\mathcal{E}}c/ZeB_{\mathrm{v}}\simeq1.6\,\mathrm{cm}$
and the minimum distance from the trap axis that the ion can approach
is $r_{\min}=\left|\mathcal{P}\right|/\sqrt{2m_{\mathrm{i}}\mathcal{E}}\simeq3.5\,\mathrm{cm}$.
It can be seen that a considerable region, corresponding to transverse
momentum in the range $P_{\perp}=\sqrt{2m_{\mathrm{i}}\mathcal{E}-P_{z}^{2}}\lesssim0.6\div0.7$,
is filled with chaotic trajectories. 

Strictly speaking, the ion dynamics should be considered separately
for the chaotic and regular trajectories. In the chaotic region, the
distribution function can be considered depending only on the integrals
of motion: energy $\mathcal{E}$ and angular momentum $\mathcal{P}$.
At the same time, in the region of regular motion, these are also
supplemented by the dependence on a third conserved quantity, for
instance, the adiabatic invariant $I_{r}$. On top of that, required
is also a correct description of the transition region, separating
chaotic and regular trajectories. This issue appears to be quite complex
and needs to be addressed in an individual paper. For this reason,
in the present article, we limit ourselves to considering the simplest
case. Namely, we further assume that the region of the hot ion regular
motion has a negligible measure, while the dynamics of the hot ions
is globally chaotic and their trajectories ergodicly fill the hypersurfaces
of constants $\mathcal{E}$ and $\mathcal{P}$. Since the chaotic
behavior appears to result from the collisionless scattering on the
non-paraxial end regions: $\left|dr_{0}/dz\right|\sim1$, we also assume
that the characteristic ergodization time is on the order of several periods
of longitudinal oscillations $\tau_{\mathrm{b}}\sim L/\sqrt{\mathcal{E}/m_{\mathrm{i}}}$.

\subsection*{Hot ion distribution function}

As noted above, we consider the hot ion density to be small enough
to neglect their collisions with each other. In addition, we neglect
the angular scattering of the hot ions in collisions with the warm
plasma particles and take into account only the warm electron drag
force
\begin{equation}
\boldsymbol{F}_{\mathrm{s}}\simeq-\nu_{\mathrm{ie}}m_{\mathrm{i}}\boldsymbol{v},\label{eq:friction_force}
\end{equation}
where $\boldsymbol{v}$ is the hot ion velocity,
\begin{equation}
\nu_{\mathrm{ie}}=1.6\times10^{-9}\mu_{\mathrm{i}}^{-1}Z^{3}\Lambda_{\mathrm{ie}}\frac{n_{\mathrm{i}}}{{\rm cm}^{-3}}\left(\frac{T}{{\rm eV}}\right)^{-\nicefrac{3}{2}}{\rm sec}^{-1},\label{eq:nu_ie}
\end{equation}
is the inverse ion slowing down time \citep{Trubnikov1965}, $\mu_{\mathrm{i}}=m_{\mathrm{i}}/m_{\mathrm{p}}$
is the ion mass $m_{\mathrm{i}}$ normalized to proton mass $m_{\mathrm{p}}$,
$\Lambda_{\mathrm{ie}}$ is the ion-electron Coulomb logarithm. In
what follows, we assume that the neutral beam is injected into the
absolute confinement region (\ref{eq:absolute_confinement}); in this
case, the loss of the hot ions can be ignored. Therefore, in the stationary
case under consideration, the kinetic equation for the distribution
function of the hot ions $f_{\mathrm{h}}$ has the following invariant
form:
\begin{equation}
\nabla_{\boldsymbol{X}}\cdot\left(\dot{\boldsymbol{X}}f_{\mathrm{h}}\right)=g_{\mathrm{h}},\label{eq:kinet_exact}
\end{equation}
where $\boldsymbol{X}$ is the six-dimensional vector of generalized
phase variables, $\dot{\boldsymbol{X}}$ is the corresponding phase
velocity vector, $\nabla_{\boldsymbol{X}}$ is the nabla differential
operator in the $\boldsymbol{X}$-space, and $g_{\mathrm{h}}=g_{\mathrm{h}}\left(\boldsymbol{X}\right)$
is the phase density of the hot ion source intensity, the explicit
form of which is determined by the injection. Specifying the variables
as $\boldsymbol{X}=\left(r,\theta,z,P_{r},P_{\theta},P_{z}\right)$,
we obtain the equations of motion in the form:
\begin{align*}
\dot{r} & =\frac{P_{r}}{m_{\mathrm{i}}}, & \dot{P}_{r} & =\frac{P_{\theta}-\Psi}{m_{\mathrm{i}}r^{2}}\left(\frac{P_{\theta}-\Psi}{r}+\partial_{r}\Psi\right)-\nu_{\mathrm{ie}}P_{r},\\
\dot{\theta} & =\frac{P_{\theta}-\Psi}{m_{\mathrm{i}}r^{2}}, & \dot{P}_{\theta} & =-\nu_{\mathrm{ie}}\left(P_{\theta}-\Psi\right),\\
\dot{z} & =\frac{P_{z}}{m_{\mathrm{i}}}, & \dot{P}_{z} & =\frac{P_{\theta}-\Psi}{m_{\mathrm{i}}r^{2}}\partial_{z}\Psi-\nu_{\mathrm{ie}}P_{z}.
\end{align*}
Since we assume axial symmetry, the distribution function $f_{\mathrm{h}}$
and the magnetic flux $\Psi$ do not depend on the cyclic variable
$\theta$. 

As mentioned above, we consider that the dynamics of ions is globally
chaotic, and the ergodization time is on the order of several periods
of longitudinal oscillations of the ion $\tau_{\mathrm{b}}\sim L/\sqrt{\mathcal{E}/m_{\mathrm{i}}}$.
Since the characteristic slowing down time $\nu_{\mathrm{ie}}^{-1}$
significantly exceeds the characteristic bounce time $\tau_{\mathrm{b}}$
(typically $\nu_{\mathrm{ie}}\tau_{\mathrm{b}}\sim10^{-3}\div10^{-4}$),
the drag force (\ref{eq:friction_force}) can be considered weak on
time scales of the period of longitudinal oscillations $\tau_{\mathrm{b}}$.
By means of the conventionally used Krylov-Bogoliubov-Mitropolsky
averaging approach \citep{Bogoliubov1961}, the slow phase space
diffusion associated with the drag force (\ref{eq:friction_force})
can be separated from the fast longitudinal oscillations. Based on
this, we replace the exact kinetic equation (\ref{eq:kinet_exact})
by an averaged one, assuming that the averaging should be performed
over the $\mathcal{E},\mathcal{P}={\rm const}$ surface in the phase
space. 

Instead of $\left(P_{r},P_{\theta},P_{z}\right)$ it is convenient
to use the new variables $\left(\mathcal{E},\mathcal{P},\alpha\right)$,
where $\alpha\in\left[0,2\pi\right)$ is defined by
\[
P_{r}=\sqrt{2m_{\mathrm{i}}\left(\mathcal{E}-\varphi_{\mathrm{eff}}\right)}\sin\alpha,\quad P_{z}=\sqrt{2m_{\mathrm{i}}\left(\mathcal{E}-\varphi_{\mathrm{eff}}\right)}\cos\alpha.
\]
Hence, we have
\[
\frac{\partial\left(P_{r},P_{\theta},P_{z}\right)}{\partial\left(\mathcal{E},\mathcal{P},\alpha\right)}=m_{\mathrm{i}},
\]
and we should also keep in mind the additional condition $\mathcal{E}>\varphi_{\mathrm{eff}}$,
which defines the region of permissible values of the new variables.
Then averaging of some quantity $\mathcal{Q}$ over the surface $\mathcal{E},\mathcal{P}={\rm const}$
is performed as follows:

\begin{equation}
\left\langle \mathcal{Q}\right\rangle _{\Gamma}\overset{\mathrm{def}}{=}\Gamma^{-1}\idotsint\limits _{\mathcal{E}>\varphi_{\mathrm{eff}}}\mathcal{Q}m_{\mathrm{i}}drd\theta dzd\alpha,\label{eq:averaging}
\end{equation}
where 
\[
\Gamma=\idotsint\limits _{\mathcal{E}>\varphi_{\mathrm{eff}}}m_{\mathrm{i}}drd\theta dzd\alpha=4\pi^{2}m_{\mathrm{i}}\iint\limits _{\mathcal{E}>\varphi_{\mathrm{eff}}}drdz,
\]
is the phase volume of the hypersurface $\mathcal{E},\mathcal{P}=\mathrm{const}$.
Finally, in the leading order with respect to $\nu_{\mathrm{ie}}\tau_{\mathrm{b}}\ll1$,
when the distribution function $f_{\mathrm{h}}$ approximately depends
only on $\mathcal{E}$ and $\mathcal{P}$, averaging the perturbed
system yields
\begin{gather}
\Gamma^{-1}\partial_{\mathcal{E}}\left(\Gamma\left\langle \dot{\mathcal{E}}\right\rangle _{\Gamma}f_{\mathrm{h}}\right)+\Gamma^{-1}\partial_{\mathcal{P}}\left(\Gamma\left\langle \dot{\mathcal{P}}\right\rangle _{\Gamma}f_{\mathrm{h}}\right)=\left\langle g_{\mathrm{h}}\right\rangle _{\Gamma},\label{eq:kinet_avereged}\\
\left\langle \dot{\mathcal{E}}\right\rangle _{\Gamma}=-2\left\langle \nu_{\mathrm{ie}}\right\rangle _{\Gamma}\mathcal{E},\quad\left\langle \dot{\mathcal{P}}\right\rangle _{\Gamma}=-\left\langle \nu_{\mathrm{ie}}\right\rangle _{\Gamma}\mathcal{P}+\left\langle \nu_{\mathrm{ie}}\Psi\right\rangle _{\Gamma}.\nonumber \\
\left\langle \dot{r}\right\rangle _{\Gamma}=0,\quad\left\langle \dot{z}\right\rangle _{\Gamma}=0,\quad\left\langle \dot{\alpha}\right\rangle _{\Gamma}=0.\nonumber 
\end{gather}

Solving the kinetic equation (\ref{eq:kinet_avereged}) one can find
the averaged distribution function of the hot ions $f_{\mathrm{h}}\simeq f_{\mathrm{h}}\left(\mathcal{E},\mathcal{P}\right)$.
This in turn enables the density of some quantity $\mathcal{Q}$ to
be obtained:
\[
\left\langle \mathcal{Q}\right\rangle _{\mathrm{h}}\overset{\mathrm{def}}{=}\frac{1}{r}\iiint\mathcal{Q}f_{\mathrm{h}}dP_{r}dP_{\theta}dP_{z}=\frac{m_{\mathrm{i}}}{r}\int\limits _{0}^{2\pi}d\alpha\int\limits _{-\infty}^{+\infty}d\mathcal{P}\int\limits _{\varphi_{\mathrm{eff}}}^{+\infty}d\mathcal{E}\mathcal{Q}f_{\mathrm{h}}.
\]
In particular, the azimuthal electric current density of the hot ions
is given by
\begin{equation}
J_{\mathrm{h}\theta}=\left\langle Zev_{\theta}\right\rangle _{\mathrm{h}}=\frac{2\pi Ze}{r^{2}}\int\limits _{-\infty}^{+\infty}d\mathcal{P}\int\limits _{\varphi_{\mathrm{eff}}}^{+\infty}d\mathcal{E}\left(\mathcal{P}-\Psi\right)f_{\mathrm{h}},\label{eq:J_h_theta}
\end{equation}
the power density of heating the warm plasma by the hot ions is defined
as
\begin{equation}
q_{\mathrm{h}}=\left\langle 2\left\langle \nu_{\mathrm{ie}}\right\rangle _{\Gamma}\mathcal{E}\right\rangle _{\mathrm{h}}=\frac{4\pi m_{\mathrm{i}}}{r}\int\limits _{-\infty}^{+\infty}d\mathcal{P}\int\limits _{\varphi_{\mathrm{eff}}}^{+\infty}d\mathcal{E}\left\langle \nu_{\mathrm{ie}}\right\rangle _{\Gamma}\mathcal{E}f_{\mathrm{h}}.\label{eq:q_h}
\end{equation}

\section{Cylindrical bubble model\label{sec:cylindrical_bubble_model}}

The complete system of equations describing the equilibrium of the
diamagnetic bubble with neutral beam injection consists of the equation
for the magnetic field (\ref{eq:grad_shafranov}) and the equations
of material equilibrium (\ref{eq:flow_equilibrium}) and thermal equilibrium
(\ref{eq:thermal_equilibrium}) for the warm plasma. This system should
be supplemented with the equation of state for the warm plasma (\ref{eq:state})
and the expression for the warm plasma current density (\ref{eq:warm_plasma_current}),
as well as the expressions for the hot ion current density (\ref{eq:J_h_theta})
and heating power density (\ref{eq:q_h}). Finally, the formulation
of the problem is completed by setting the boundary conditions for
the magnetic field: (\ref{eq:field_external}), (\ref{eq:field_internal}),
and (\ref{eq:core_radius}), the latter of which determines the boundary
of the bubble core, and also the internal (\ref{eq:flow_equilibrium_internal}),
(\ref{eq:thermal_equilibrium_internal}), and external (\ref{eq:equilibrium_external})
conditions for the warm plasma equilibrium equations.

Just as it was done by \citet{Beklemishev2016} for the case of MHD
equilibrium, we further consider a simplified model of a cylindrical
diamagnetic bubble. In the central part of such a bubble, there is
a cylindrical core of length $L$ and radius $a$ with zero magnetic
field: $\psi\equiv0$; outside the core, for $r>a$, the magnetic
field lines are straight: $\psi=\psi\left(r\right)$. At the ends
of the central cylindrical part, there are non-paraxial regions gradually
turning into the mirror throats with the magnetic field $B_{\mathrm{m}}\simeq\mathrm{const}$.
In addition, the source of the warm plasma is further considered to
be entirely contained inside the bubble core. In this case, the density
of the warm plasma inside the core is expected to be much higher than
outside. Since the drag force (\ref{eq:friction_force}) is proportional
to the density of the warm plasma, the hot ions seem to slow down
and release the energy mainly inside the bubble core as well. 

\subsection{Magnetic field distribution}

The equilibrium equation for the magnetic field (\ref{eq:grad_shafranov})
can only be considered in the central cylindrical region, and the
magnetic flux in the mirrors is considered to be given: $\psi_{\mathrm{m}}=B_{\mathrm{m}}\pi r^{2}$.
Then, after substituting the current density of the warm plasma (\ref{eq:warm_plasma_current}),
the equation (\ref{eq:grad_shafranov}) predictably reduces to the
paraxial equilibrium with the Lorentz force from the hot ions:
\begin{gather}
\frac{d}{d\psi}\left(\frac{B^{2}}{8\pi}+p\right)=-\frac{J_{\mathrm{h}\theta}}{2\pi rc},\label{eq:magnetic_field_CBM}\\
\frac{dr}{d\psi}=\frac{1}{2\pi r}\frac{1}{B}.\label{eq:magnetic_flux_CBM}
\end{gather}
In this equations we use the flux coordinate $\psi$, so $r=r\left(\psi\right)$
is the inverse function of $\psi=\psi\left(r\right)$ and is essentially
the radius of the magnetic surface that corresponds to the magnetic
flux $\psi$. In addition, the magnetic field $B$ and the current
density of the hot ions $J_{\mathrm{h}\theta}$ should also be considered
here as functions of $\psi$. The corresponding boundary conditions
(\ref{eq:field_external}), (\ref{eq:field_internal}), and (\ref{eq:core_radius})
can be reduced to
\begin{gather}
\left.B\right|_{\psi=0}=0,\quad\left.r\right|_{\psi=0}=a,\label{eq:magnetic_field_CBM_BC_0}\\
\left.B\right|_{\psi\rightarrow+\infty}=B_{\mathrm{v}},\label{eq:magnetic_field_CBM_BC_out}
\end{gather}
where $B_{\mathrm{v}}=\mathrm{const}$ is the vacuum magnetic field
at the central section of the trap.

\subsection{Warm plasma equilibrium}

Warm plasma equilibrium equations (\ref{eq:flow_equilibrium}) and
(\ref{eq:thermal_equilibrium}) mainly remain the same except for
the transport coefficients $\Lambda_{{\rm i}\perp}$ and $\Lambda_{{\rm E}\perp}$,
which are reduced to a simpler form in the cylindrical bubble approximation:
\begin{gather}
-\frac{d}{d\psi}\left(\Lambda_{\mathrm{i}\perp}\frac{dp}{d\psi}\right)+\frac{2u_{\mathrm{m}}}{B_{\mathrm{m}}}n_{\mathrm{i}}=0,\label{eq:equilibrium_1_CBM}\\
-\frac{d}{d\psi}\left(\frac{5}{2}\frac{p}{n_{\mathrm{i}}}\Lambda_{\mathrm{i}\perp}\frac{dp}{d\psi}\right)-\frac{d}{d\psi}\left(\Lambda_{\mathrm{E}\perp}\frac{dT}{d\psi}\right)+\frac{2\alpha_{\mathrm{E}}u_{\mathrm{m}}}{B_{\mathrm{m}}}n_{\mathrm{i}}T=0,\label{eq:equilibrium_2_CBM}\\
\Lambda_{\mathrm{i}\perp}\simeq4\pi^{2}c^{2}n_{\mathrm{i}}\frac{r^{2}L}{\sigma_{\mathrm{w}}B},\quad\Lambda_{\mathrm{E}\perp}\simeq4\pi^{2}\varkappa_{\mathrm{w}}Br^{2}L.\nonumber 
\end{gather}
Here we also take into account that due to the hot ions mainly slowing
down inside the bubble core, the external hot ion heating power release
is negligible, i.e. $Q_{\mathrm{h}}\equiv0$ and $W_{\mathrm{i}}\equiv0$
for $\psi>0$. The boundary conditions (\ref{eq:flow_equilibrium_internal}),
(\ref{eq:thermal_equilibrium_internal}), and (\ref{eq:equilibrium_external}),
in turn, remain exactly the same, except that $Q_{\mathrm{h}0}$ and
$W_{\mathrm{i}0}$ now have the meaning of the total absorbed injection
power and the total warm plasma source, respectively.

\subsection{Hot ion equilibrium\label{subsec:hot_ion_equilibrium}}

Consider the injection of a monoenergetic neutral beam with the injection
energy $\mathcal{E}_{\mathrm{NB}}$ and the total absorbed injection
power $Q_{\mathrm{h}0}$. The injection is carried out at the angle
$\xi_{\mathrm{NB}}\in\left(0,\pi/2\right)$ to the axis of the trap,
and the distance from the beam to the trap axis is finite and equal
to $r_{\mathrm{NB}}<a$. In this case, the source in the kinetic equation
(\ref{eq:kinet_avereged}) has the form\footnote{In a real experiment, the beam has a finite width and angular spread,
and it is also not exactly monoenergetic. Nevertheless, any injection
can be represented as a combination of beams of type (\ref{eq:g_h_avg}).
In other words, the solution of the kinetic equation with such a right-hand
side is essentially a Green's function.}:
\begin{equation}
\left\langle g_{\mathrm{h}}\right\rangle _{\Gamma}=\frac{1}{\Gamma}\frac{Q_{\mathrm{h}0}}{\mathcal{E}_{\mathrm{NB}}}\delta\left(\mathcal{E}-\mathcal{E}_{\mathrm{NB}}\right)\delta\left(\mathcal{P}-\mathcal{P}_{\mathrm{NB}}\right),\label{eq:g_h_avg}
\end{equation}
where $\delta\left(x\right)$ is the Dirac delta function, 
\begin{equation}
\mathcal{P}_{\mathrm{NB}}=-\sqrt{2m_{\mathrm{i}}\mathcal{E}_{\mathrm{NB}}}r_{\mathrm{NB}}\sin\xi_{\mathrm{NB}}<0\label{eq:P_NB}
\end{equation}
is the angular momentum of the injected ions. Due to the beam slowing
down mainly in the core of the bubble, averaging in (\ref{eq:kinet_avereged})
yields
\[
\left\langle \nu_{\mathrm{ie}}\right\rangle _{\Gamma}\simeq\nu_{\mathrm{ie}0},\quad\left\langle \nu_{\mathrm{ie}}\Psi\right\rangle _{\Gamma}\simeq0,
\]
where $\nu_{\mathrm{ie}0}=\left.\nu_{\mathrm{ie}}\right|_{T=T_{0},n_{\mathrm{i}}=n_{\mathrm{i}0}}$.
The resulting kinetic equation
\[
-2\nu_{\mathrm{ie}0}\partial_{\mathcal{E}}\left(\mathcal{E}\Gamma f_{\mathrm{h}}\right)-\nu_{\mathrm{ie}0}\partial_{\mathcal{P}}\left(\mathcal{P}\Gamma f_{\mathrm{h}}\right)\simeq\frac{Q_{\mathrm{h}0}}{\mathcal{E}_{\mathrm{NB}}}\delta\left(\mathcal{E}-\mathcal{E}_{\mathrm{NB}}\right)\delta\left(\mathcal{P}-\mathcal{P}_{\mathrm{NB}}\right),
\]
is satisfied by
\begin{equation}
f_{\mathrm{h}}\simeq\frac{1}{2\nu_{\mathrm{ie}0}}\frac{1}{\Gamma}\frac{Q_{\mathrm{h}0}}{\mathcal{E}_{\mathrm{NB}}}\frac{1}{\mathcal{E}^{\nicefrac{3}{2}}}\delta\left(\frac{\mathcal{P}}{\sqrt{\mathcal{E}}}-\frac{\mathcal{P}_{\mathrm{NB}}}{\sqrt{\mathcal{E}_{\mathrm{NB}}}}\right),\quad\mathcal{E}<\mathcal{E}_{\mathrm{NB}},\label{eq:f_h}
\end{equation}
where the phase volume $\Gamma$ is approximately equal to
\[
\Gamma=4\pi^{2}m_{\mathrm{i}}L\int\limits _{\mathcal{E}>\varphi_{\mathrm{eff}}}dr,
\]
and $\varphi_{\mathrm{eff}}$ is the effective potential (\ref{eq:phi_eff})
in the central plane. 

The remaining integral in $\Gamma$ can be represented in a simpler
form. To do this, we further assume that, first, there is no reversed
field and, second, the magnetic field either increases with the radius
or decreases not faster than $r^{-1}$. In other words, the following
conditions are met:
\[
B\geq0,\quad\frac{r}{B}\frac{dB}{dr}\geq-1.
\]
It can be shown that in this case, the effective potential $\varphi_{\mathrm{eff}}$
has only one minimum. Consequently, the equation $\mathcal{E}=\varphi_{\mathrm{eff}}$
has no more than two roots: $r_{\min}=r_{\min}\left(\mathcal{E},\mathcal{P}\right)$
and $r_{\max}=r_{\max}\left(\mathcal{E},\mathcal{P}\right)$, which
are the radial boundaries of the integration domain $\mathcal{E}>\varphi_{\mathrm{eff}}$.
These roots are essentially the minimum and maximum distance from
the axis available to a hot ion with energy $\mathcal{E}$ and angular
momentum $\mathcal{P}$. In what follows, $r_{\min}$ and $r_{\max}$
are shortly referred to as the \textit{minimum radius} and the \textit{maximum
radius}, respectively, and we also define the corresponding minimum
and maximum radii for the injected ions: $\overline{r}_{\min}=r_{\min}\left(\mathcal{E}_{\mathrm{NB}},\mathcal{P}_{\mathrm{NB}}\right)$
and $\overline{r}_{\max}=r_{\max}\left(\mathcal{E}_{\mathrm{NB}},\mathcal{P}_{\mathrm{NB}}\right)$.
Eventually, the phase volume can be represented in the following form:
\[
\Gamma=4\pi^{2}m_{\mathrm{i}}L\left(r_{\max}-r_{\min}\right).
\]

It is worth noting that the minimum radius for the ions passing through
the interior of the bubble core can be found explicitly. Indeed, since
$\Psi\equiv0$ in the bubble core, for $r_{\min}<a$ we have:
\[
\mathcal{E}=\varphi_{\mathrm{eff}}\left(r_{\min}\right)=\frac{\left|\mathcal{P}\right|^{2}}{2m_{\mathrm{i}}r_{\min}^{2}},\quad\Rightarrow\quad r_{\min}=\frac{\left|\mathcal{P}\right|}{\sqrt{2m_{\mathrm{i}}\mathcal{E}}}.
\]
Note that the solution (\ref{eq:f_h}) also implies that in the approximation
considered, the hot ion energy and angular momentum are related as
follows: $\mathcal{P}\sqrt{\mathcal{E}_{\mathrm{NB}}}=\mathcal{P}_{\mathrm{NB}}\sqrt{\mathcal{E}}$.
This means that the minimum radius for the ions injected in the bubble
core ($r_{\mathrm{NB}}<a$) remains constant: 
\[
r_{\min}=\overline{r}_{\min}=\frac{\left|\mathcal{P}_{\mathrm{NB}}\right|}{\sqrt{2m_{\mathrm{i}}\mathcal{E}_{\mathrm{NB}}}}.
\]
Given (\ref{eq:P_NB}), the minimum radius can also be expressed in
terms of the injection radius $r_{\mathrm{NB}}$ and injection angle
$\xi_{\mathrm{NB}}$: $\overline{r}_{\min}=r_{\mathrm{NB}}\sin\xi_{\mathrm{NB}}$.
In addition, it follows that the sign of the angular momentum does
not change during the ion slowing down, and the ions initially injected
into the absolute confinement region (\ref{eq:absolute_confinement})
always stay in it:
\[
-\mathcal{R}\Omega_{\mathrm{i}}\mathcal{P}_{\mathrm{NB}}>\mathcal{E}_{\mathrm{NB}},\quad\Rightarrow\quad-\mathcal{R}\Omega_{\mathrm{i}}\mathcal{P}>\sqrt{\mathcal{E}\mathcal{E}_{\mathrm{NB}}}\geq\mathcal{E}.
\]
Therefore, the axial loss of the hot ions can indeed be neglected\footnote{In fact, collisional angular scattering at low energies $\mathcal{E}\sim T$
of course should eventually lead to the loss of the hot ions. The
balance of the hot ions with the distribution (\ref{eq:f_h}) essentially
assumes the presence of a sink in phase space at $\mathcal{E}\rightarrow0$.}.

Eventually, using the resulting distribution function (\ref{eq:f_h})
and the definition (\ref{eq:J_h_theta}), we obtain the azimuthal
diamagnetic current density of the hot ions:
\begin{equation}
\frac{J_{\mathrm{h}\theta}}{2\pi rc}=-\frac{\overline{r}_{\min}}{a}\frac{\Pi_{\mathrm{h}}}{\psi_{\mathrm{h}}}\frac{a^{3}}{r^{3}}\Theta\left(\frac{r-\overline{r}_{\min}}{a-\overline{r}_{\min}}-\frac{\psi}{\psi_{\mathrm{h}}}\right)\int\limits _{\frac{a-\overline{r}_{\min}}{r-\overline{r}_{\min}}\frac{\psi}{\psi_{\mathrm{h}}}}^{1}\frac{a-\overline{r}_{\min}}{r_{\max}^{*}\left(\eta\right)-\overline{r}_{\min}}\left(1+\frac{a-\overline{r}_{\min}}{\overline{r}_{\min}}\frac{\psi}{\psi_{\mathrm{h}}}\frac{1}{\eta}\right)d\eta,\label{eq:J_h_theta_CBM}
\end{equation}
where $\psi_{\mathrm{h}}=2\pi B_{\mathrm{v}}\left(a-\overline{r}_{\min}\right)\rho_{\mathrm{NB}}$
is the characteristic magnetic flux induced by the hot ion current,
$\rho_{\mathrm{NB}}=\sqrt{2m_{\mathrm{i}}\mathcal{E}_{\mathrm{NB}}}c/ZeB_{\mathrm{v}}$
is the characteristic Larmor radius of injected ions, the coefficient
\[
\Pi_{\mathrm{h}}=\frac{Q_{\mathrm{h}0}}{\nu_{\mathrm{ie}0}\pi a^{2}L}
\]
is the characteristic energy density of the hot ions, and
\[
r_{\max}^{*}\left(\eta\right)\overset{\mathrm{def}}{=}r_{\max}\left(\mathcal{E}_{\mathrm{NB}}\eta^{2},\mathcal{P}_{\mathrm{NB}}\eta\right)
\]
is the maximum radius of the hot ions with the distribution (\ref{eq:f_h}),
which is to be found from the equation $\mathcal{E}=\varphi_{\mathrm{eff}}$
reduced to explicit form:
\begin{equation}
r_{\max}^{*}\left(\eta\right)=\left.r\left(\psi\right)\right|_{\psi=\psi_{\mathrm{h}}\eta\frac{r_{\max}^{*}\left(\eta\right)-\overline{r}_{\min}}{a-\overline{r}_{\min}}}.\label{eq:r_max_asterisk}
\end{equation}
In particular, the outer boundary is $\overline{r}_{\max}=r_{\max}^{*}\left(1\right)$.

\section{Thin transition layer limit\label{sec:thin_transition_layer_limit}}

The resulting complete system of equilibrium equations for a cylindrical
bubble (\ref{eq:magnetic_field_CBM}), (\ref{eq:magnetic_flux_CBM}),
(\ref{eq:equilibrium_1_CBM}), and (\ref{eq:equilibrium_2_CBM}) accompanied
by the expression for the hot ion current (\ref{eq:J_h_theta_CBM})
proves to be essentially nonlinear. An exact equilibrium solution
can only be obtained numerically. Detailed numerical simulations and
analysis of the corresponding numerical equilibria are planned to
be provided in future work. In the present paper, we focus on considering
a limiting case that allows a significant simplification of the equations.

As mentioned at the beginning of Section \ref{sec:basic_assumptions_of_theoretical_model},
at the boundary of the diamagnetic bubble, right beyond the core,
there is a transition layer inside which the magnetic field changes
from $B=0$ to $B=B_{\mathrm{v}}$. The total thickness of this layer
– the radial size of the region such that $0<B<B_{\mathrm{v}}$ –
is further denoted by $\lambda$. The inner boundary of the layer
$r=a$ corresponds to the radial boundary of the bubble core, where
the magnetic field is zero: $\left.B\right|_{r\leq a}=0$. The outer
boundary $r=a+\lambda$ is the radius at which the diamagnetic current
density of the plasma vanishes and the magnetic field reaches the
vacuum value: $\left.B\right|_{r\geq a+\lambda}=B_{\mathrm{v}}$.
In other words, the outer boundary of the transition layer represents
the boundary of the plasma. In the same way, we define separately
the boundaries for the warm plasma $r=a+\lambda_{\mathrm{w}}$ and
hot ion ions $r=a+\lambda_{\mathrm{h}}$, beyond which the corresponding
diamagnetic currents vanish; $\lambda_{\mathrm{w}}$ and $\lambda_{\mathrm{h}}$
are further naturally called the thicknesses of the transition layer
for the warm plasma and the hot ions, respectively. It is clear that
the total thickness of the transition layer is determined by the largest
one: $\lambda=\max\left\{ \lambda_{\mathrm{w}},\lambda_{\mathrm{h}}\right\} $. 

The transition layer thickness for the warm plasma $\lambda_{\mathrm{w}}$
is determined by the characteristic scale of the warm plasma resistive
transverse diffusion across the magnetic field, which is normally
quite small. In particular, the MHD equilibrium model \citep{Beklemishev2016}
shows that the thickness of the transition layer proves to be approximately
equal to $\lambda_{\mathrm{w}\,\mathrm{MHD}}=7\lambda_{\mathrm{GD}}$,
where
\begin{equation}
\lambda_{\mathrm{GD}}=\sqrt{\frac{c^{2}}{4\pi\sigma_{\mathrm{w}}}\tau_{\mathrm{GD}}}\label{eq:lambda_GD}
\end{equation}
is the characteristic thickness of magnetic field diffusion into the
plasma and $\tau_{\mathrm{GD}}=\mathcal{R}L/2u_{\mathrm{m}}$ is the
gas-dynamic lifetime. In the case of 'classical' Spitzer conductivity
$\sigma_{\mathrm{w}}=Z^{2}e^{2}n_{\mathrm{i}}/m_{\mathrm{i}}\nu_{\mathrm{ie}}$,
for the typical parameters: $T\sim1\,\mathrm{keV}$, $\mathcal{R}\sim15$,
$L\sim0.5\,\mathrm{cm}$, the quantity $\lambda_{\mathrm{GD}}$ is
on the order of tenths of a centimeter. However, in the presence of
the hot ion component, MHD is not applicable, and this estimate ceases
to be valid. 

For the distribution (\ref{eq:f_h}), the outer boundary, at which
the current of the hot ions (\ref{eq:J_h_theta_CBM}) vanishes, corresponds
to the maximum radius of the injected ions $\overline{r}_{\max}$.
Then the thickness of the hot ion transition layer in this case is
equal to $\lambda_{\mathrm{h}}=\overline{r}_{\max}-a$. On the other
hand, the hot ion transition layer is essentially the boundary region
inside which the hot ions are 'reflected' from the external vacuum
magnetic field outside the bubble core, and its thickness seems to
be proportional to the Larmor radius of the injected ions $\lambda_{\mathrm{h}}\sim\rho_{\mathrm{NB}}$.
Under typical conditions, the hot ion Larmor radius $\rho_{\mathrm{NB}}$
is on the order of centimeters and normally proves to be much greater
than the warm plasma transverse diffusion scale $\lambda_{\mathrm{GD}}$.
Thus, the total transition layer thickness $\lambda=\max\left\{ \lambda_{\mathrm{w}},\lambda_{\mathrm{h}}\right\} $
could be expected mainly determined by the hot ion transition layer
thickness: $\lambda_{\mathrm{h}}\gg\lambda_{\mathrm{w}}$.

Further in this work, we consider the limit of the thin transition
layer $\lambda\ll a$, which allows the model equations to be greatly
simplified. Since $\lambda$ normally decreases with increasing magnetic
field, this approximation seems to correspond to the case of a sufficiently
strong vacuum field $B_{\mathrm{v}}$.

\subsection{Equilibrium magnetic field distribution\label{subsec:equilibrium_magnetic_field}}

The magnetic field equilibrium is determined by the equations (\ref{eq:magnetic_field_CBM})
and (\ref{eq:magnetic_flux_CBM}) with the boundary conditions (\ref{eq:magnetic_field_CBM_BC_0})
and (\ref{eq:magnetic_field_CBM_BC_out}). When solving these equations,
we consider the pressure profile of the warm plasma $p=p\left(\psi\right)$
given, implying that it can be found from the corresponding equilibrium
equations (\ref{eq:equilibrium_1_CBM}) and (\ref{eq:equilibrium_2_CBM}).
At the same time, the diamagnetic current of the hot ions is determined
by the expression (\ref{eq:J_h_theta_CBM}). 

It is further convenient to use the following dimensionless quantities:
\[
\phi=\frac{\psi}{\psi_{\mathrm{h}}},\quad x=\frac{r}{a},\quad R=\frac{B}{B_{\mathrm{v}}},\quad\beta_{\mathrm{w}}=\frac{8\pi p}{B_{\mathrm{v}}^{2}},
\]
where $\beta_{\mathrm{w}}$ is the warm plasma pressure normalized
to the energy density of the vacuum magnetic field $B_{\mathrm{v}}^{2}/8\pi$,
also referred to simply as the relative pressure of the warm plasma.
According to this, we also define the dimensionless minimum and maximum
radii as follows:
\[
\overline{x}_{\min}=\frac{\overline{r}_{\min}}{a},\quad x_{\max}^{*}\left(\eta\right)=\frac{r_{\max}^{*}\left(\eta\right)}{a},
\]
where the latter is determined by the equation
\begin{equation}
x_{\max}^{*}\left(\eta\right)=\left.x\left(\phi\right)\right|_{\phi=\eta\frac{x_{\max}^{*}\left(\eta\right)-\overline{x}_{\min}}{1-\overline{x}_{\min}}},\label{eq:x_max_asterisk}
\end{equation}
which results from the corresponding normalization of the equation
(\ref{eq:r_max_asterisk}). Then the equations (\ref{eq:magnetic_field_CBM})
and (\ref{eq:magnetic_flux_CBM}) with substituted hot ion current
(\ref{eq:J_h_theta_CBM}) take the following dimensionless form:
\begin{gather}
\frac{d}{d\phi}\left(R^{2}+\beta_{\mathrm{w}}\right)=\frac{\Lambda_{\mathrm{h}}}{x^{3}}\Theta\left(\frac{x-\overline{x}_{\min}}{1-\overline{x}_{\min}}-\phi\right)\int\limits _{\frac{1-\overline{x}_{\min}}{x-\overline{x}_{\min}}\phi}^{1}\frac{1-\overline{x}_{\min}}{x_{\max}^{*}\left(\eta\right)-\overline{x}_{\min}}\left(1+\frac{1-\overline{x}_{\min}}{\overline{x}_{\min}}\frac{\phi}{\eta}\right)d\eta,\label{eq:magnetic_flux_CBM_norm}\\
\frac{dx}{d\phi}=\frac{\epsilon}{x}\frac{1}{R},\label{eq:magnetic_field_CBM_norm}
\end{gather}
where
\[
\Lambda_{\mathrm{h}}=\overline{x}_{\min}\frac{8\pi\Pi_{\mathrm{h}}}{B_{\mathrm{v}}^{2}},\quad\epsilon=\left(1-\overline{x}_{\min}\right)\frac{\rho_{\mathrm{NB}}}{a}.
\]
Normalizing the boundary conditions (\ref{eq:magnetic_field_CBM_BC_0})
and (\ref{eq:magnetic_field_CBM_BC_out}) yields:
\begin{gather}
\left.R\right|_{\phi=0}=0,\quad\left.x\right|_{\phi=0}=1,\label{eq:magnetic_field_CBM_BC_norm_0}\\
\left.R\right|_{\phi\rightarrow+\infty}=1.\label{eq:magnetic_field_CBM_BC_norm_out}
\end{gather}

As mentioned above, the hot ion transition layer thickness $\lambda_{\mathrm{h}}$
appears to be on the order of the Larmor radius of the injected ions
$\rho_{\mathrm{NB}}$, which means that in the thin transition layer
limit before us $\lambda_{\mathrm{h}}\ll a$, we should also expect
the ratio $\rho_{\mathrm{NB}}/a$ to be small. Given this, it seems
appropriate to apply the method of successive approximations by considering
the coefficient $\epsilon\propto\rho_{\mathrm{NB}}/a$ in the right-hand
side of the equation (\ref{eq:magnetic_field_CBM_norm}) as a small
expansion parameter. Namely, we further assume that the solution of
the system (\ref{eq:magnetic_flux_CBM_norm}) and (\ref{eq:magnetic_field_CBM_norm})
can be represented in the form of an asymptotic expansion.

For a given magnetic field profile $R=R\left(\phi\right)$, by integrating
the equation (\ref{eq:magnetic_field_CBM_norm}) and taking into account
the boundary condition (\ref{eq:magnetic_field_CBM_BC_norm_0}), the
radius as a function of the magnetic flux can be explicitly expressed:
\begin{equation}
x\left(\phi\right)=\sqrt{1+2\epsilon\int\limits _{0}^{\phi}\frac{d\phi^{\prime}}{R\left(\phi^{\prime}\right)}}.\label{eq:x_phi}
\end{equation}
In the leading order of the approximation $\epsilon\ll1$, when small
corrections are completely neglected, the expression (\ref{eq:x_phi})
reduces to

\begin{equation}
x\left(\phi\right)\simeq1+o\left(\epsilon^{0}\right).\label{eq:x_phi_0}
\end{equation}
Substituting (\ref{eq:x_phi_0}) into the equation (\ref{eq:x_max_asterisk})
yields the corresponding approximation for the maximum radius:
\begin{equation}
x_{\max}^{*}\left(\eta\right)\simeq1+o\left(\epsilon^{0}\right).\label{eq:x_max_asterisk_0}
\end{equation}
The obtained formulas (\ref{eq:x_phi_0}) and (\ref{eq:x_max_asterisk})
correspond to $r\simeq a$. In other words, on the scale of the transition
layer, the radius $r$ can be considered almost constant and equal
to the bubble radius $a$, while the magnetic flux $\psi$ varies
greatly. At a qualitative level, this is associated with the high
density of the diamagnetic current inside the layer. 

Taking into account (\ref{eq:x_phi_0}) and (\ref{eq:x_max_asterisk_0}),
the integral in the right-hand side of the equation (\ref{eq:magnetic_flux_CBM_norm})
is evaluated explicitly:
\begin{multline*}
\frac{d}{d\phi}\left(R^{2}+\beta_{\mathrm{w}}\right)\simeq\Lambda_{\mathrm{h}}\Theta\left(1-\phi\right)\int\limits _{\phi}^{1}\left[1+\left(\overline{x}_{\min}^{-1}-1\right)\frac{\phi}{\eta}\right]d\eta+o\left(\epsilon^{0}\right)=\\
=\Lambda_{\mathrm{h}}\left[1-\phi-\left(\overline{x}_{\min}^{-1}-1\right)\phi\ln\phi\right]\Theta\left(1-\phi\right)+o\left(\epsilon^{0}\right).
\end{multline*}
As can be seen, the outer boundary of the hot ions in this order of
approximation corresponds to $\phi=1$. Given the boundary conditions
(\ref{eq:magnetic_field_CBM_BC_norm_0}) and (\ref{eq:magnetic_field_CBM_BC_norm_out}),
the resulting equation yields:
\begin{equation}
R\left(\phi\right)\simeq R^{\left(0\right)}\left(\phi\right)+o\left(\epsilon^{0}\right),\label{eq:R_phi_0}
\end{equation}
where
\[
R^{\left(0\right)}\left(\phi\right)=\begin{cases}
\sqrt{\beta_{\mathrm{w}0}-\beta_{\mathrm{w}}\left(\phi\right)+\Lambda_{\mathrm{h}}\phi\left[1-\dfrac{\phi}{2}-\left(\overline{x}_{\min}^{-1}-1\right)\dfrac{\phi}{2}\left(\ln\phi-\dfrac{1}{2}\right)\right]}, & \phi<1,\\
\sqrt{1-\beta_{\mathrm{w}}\left(\phi\right)}, & \phi\geq1.
\end{cases}
\]
and $\beta_{\mathrm{w}0}=\beta_{\mathrm{w}}\left(0\right)=8\pi p_{0}/B_{\mathrm{v}}^{2}$
is the relative pressure of the warm plasma inside the bubble core.
In the absence of surface currents, the solution (\ref{eq:R_phi_0})
should be continuous at the boundary $\phi=1$, from which we also
obtain the following condition:
\begin{equation}
1\simeq\beta_{\mathrm{w}0}+\beta_{\mathrm{h}0}+o\left(\epsilon^{0}\right),\label{eq:force_balance_0}
\end{equation}
Here, the quantity
\[
\beta_{\mathrm{h}0}=\frac{1}{4}\left(\overline{x}_{\min}^{-1}+1\right)\Lambda_{\mathrm{h}}=\left(1+\overline{x}_{\min}\right)\frac{2\pi\Pi_{\mathrm{h}}}{B_{\mathrm{v}}^{2}}
\]
can be interpreted as the characteristic relative energy density of
the hot ions in the bubble. The formula (\ref{eq:force_balance_0})
essentially expresses the balance between the plasma pressure inside
the bubble and the pressure of the external magnetic field.

By applying the successive approximations to find higher orders of
the expansion, the solution can be further refined. In particular,
taking into account first order corrections in the expression (\ref{eq:x_phi})
yields:
\[
x\left(\phi\right)\simeq1+\epsilon\int\limits _{0}^{\phi}\frac{d\phi^{\prime}}{R^{\left(0\right)}\left(\phi^{\prime}\right)}+o\left(\epsilon^{1}\right).
\]
Using the obtained dependence, the corresponding correction to the
maximum radius is also found from the equation (\ref{eq:x_max_asterisk}):
\[
x_{\max}^{*}\left(\eta\right)\simeq1+\epsilon\int\limits _{0}^{\eta}\frac{d\phi}{R^{\left(0\right)}\left(\phi\right)}+o\left(\epsilon^{1}\right).
\]
For $\eta=1$ this formula determines the thickness of the hot ion
transition layer:
\begin{equation}
\frac{\lambda_{\mathrm{h}}}{a}=x_{\max}^{*}\left(1\right)-1\simeq\epsilon\int\limits _{0}^{1}\frac{d\phi}{R^{\left(0\right)}\left(\phi\right)}+o\left(\epsilon^{1}\right).\label{eq:lambda_h}
\end{equation}

\begin{figure}
\begin{centering}
\includegraphics{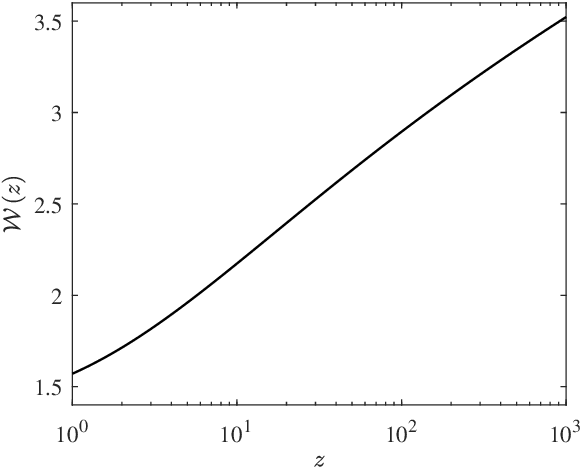}
\par\end{centering}
\caption{Function $\mathcal{W}=\mathcal{W}\left(z\right)$ defined by the expression
(\ref{eq:w_z}). \label{fig:w_z}}
\end{figure}

Further in this paper, we consider the equilibrium configuration of
the magnetic field to be approximately defined by (\ref{eq:x_phi_0})
and (\ref{eq:R_phi_0}) at $\epsilon\rightarrow0$. In other words,
we take into account only the leading order of the asymptotic expansion.
The applicability criterion for this approximation is determined by
the limit
\[
\frac{1-x_{\max}^{*}\left(1\right)}{1-\overline{x}_{\min}}=\frac{\lambda_{\mathrm{h}}}{a-\overline{r}_{\min}}\ll1,
\]
which is actually assumed in the derivation of (\ref{eq:R_phi_0}).
After substituting (\ref{eq:lambda_h}), we arrive at the following
condition:
\[
\int\limits _{0}^{1}\frac{d\phi}{R^{\left(0\right)}\left(\phi\right)}\ll\frac{a}{\rho_{\mathrm{NB}}}.
\]
An upper bound for the integral involved can be obtained as follows:
\[
\int\limits _{0}^{1}\frac{d\phi}{R^{\left(0\right)}\left(\phi\right)}\leq\frac{1}{\sqrt{1-\beta_{\mathrm{w}0}}}\mathcal{W}\left(\frac{a}{\overline{r}_{\min}}\right),
\]
where
\begin{equation}
\mathcal{W}\left(z\right)\overset{\mathrm{def}}{=}\frac{\sqrt{z+1}}{2}\int\limits _{0}^{1}\frac{d\phi}{\sqrt{\phi\left(1-\frac{\phi}{2}-\left(z-1\right)\frac{\phi}{2}\left(\ln\phi-\frac{1}{2}\right)\right)}}.\label{eq:w_z}
\end{equation}
As can be seen from Figure \ref{fig:w_z}, for all reasonable $a/\overline{r}_{\min}$
the function $\mathcal{W}$ is on the order of $2\div3$.

Before moving on to considering the equilibrium of the warm plasma,
it is also useful to examine the expression (\ref{eq:force_balance_0})
in greater detail. For given parameters of the warm plasma inside
the bubble core: density $n_{\mathrm{i}0}$ and temperature $T_{0}$,
which are to be found from the solution of the equilibrium equations
(\ref{eq:equilibrium_1_CBM}) and (\ref{eq:equilibrium_2_CBM}), the
expression (\ref{eq:force_balance_0}) defines the relation between
the radius of the core $a$ and the fixed external parameters: absorbed
heating power $Q_{\mathrm{h}0}$, vacuum magnetic field $B_{\mathrm{v}}$,
and geometric parameters $L$ and $\overline{r}_{\min}$. After expressing
the warm plasma density $n_{\mathrm{i}0}$ from the equation of state
(\ref{eq:state}), the formula (\ref{eq:force_balance_0}) is reduced
to a cubic equation for the normalized core radius $k=a/\overline{r}_{\min}=\overline{x}_{\min}^{-1}$:
\begin{equation}
k^{3}-\left(4\tau_{\mathrm{s}}\nu_{\mathrm{ie}0}\right)^{-1}\left(k+1\right)\simeq0,\label{eq:a}
\end{equation}
where
\[
\tau_{\mathrm{s}}=\beta_{\mathrm{h}0}\frac{B_{\mathrm{v}}^{2}}{8\pi}\frac{\pi\overline{r}_{\min}^{2}L}{Q_{\mathrm{h}0}}.
\]
Provided that the core radius $a$ exceeds the minimum radius $\overline{r}_{\min}$,
i.e. $k\geq1$, this equation has exactly one real root for a fixed
$\beta_{\mathrm{w}0}\in\left[0,1\right]$:
\[
k = \begin{cases}
\dfrac{3}{\mathfrak{D}}\cos\left(\dfrac{1}{3}\arccos\mathfrak{D}\right), & 0<\mathfrak{D}\leq1,\\
\dfrac{3}{\mathfrak{D}}\cosh\left(\dfrac{1}{3}\mathrm{arccosh}\,\mathfrak{D}\right), & 1<\mathfrak{D},
\end{cases}
\]
where $\mathfrak{D}=\sqrt{27\tau_{\mathrm{s}}\nu_{\mathrm{ie}0}}$
and $\mathrm{arccosh}\,x=\ln\left(x+\sqrt{x^{2}-1}\right)$. This means
that for a given temperature in the core $T_{0}$, which we further
consider determined by (\ref{eq:T_0}), this root specifies the functional
dependency of the core radius on the pressure in the core: $k=k\left(\beta_{\mathrm{w}0}\right)$.

\subsection{Warm plasma equilibrium\label{subsec:warm_plasma_equilibrium}}

The equilibrium of the warm plasma is described by the equations (\ref{eq:equilibrium_1_CBM})
and (\ref{eq:equilibrium_2_CBM}) with the boundary conditions (\ref{eq:flow_equilibrium_internal}),
(\ref{eq:thermal_equilibrium_internal}) and (\ref{eq:equilibrium_external}).
In what follows, the electrical and thermal conductivity of the warm
plasma are considered classical \citep{Braginskii1965}:
\begin{gather}
\sigma_{\mathrm{w}}=\frac{Z^{2}e^{2}n_{\mathrm{i}}}{m_{\mathrm{i}}\nu_{\mathrm{ie}}}=8.7\times10^{13}Z^{-1}\Lambda_{\mathrm{ie}}^{-1}\left(\frac{T}{\mathrm{eV}}\right)^{\nicefrac{3}{2}}\mathrm{sec}^{-1},\label{eq:Spitzer}\\
\varkappa_{\mathrm{w}}=10^{-3}\left(\mu_{\mathrm{i}}^{\nicefrac{1}{2}}+0.077Z\right)Z^{2}\Lambda_{\mathrm{ie}}\left(\frac{B}{\mathrm{G}}\right)^{-2}\left(\frac{n_{\mathrm{i}}}{{\rm cm}^{-3}}\right)^{2}\left(\frac{T}{{\rm eV}}\right)^{-\nicefrac{1}{2}}{\rm cm}^{-1}{\rm sec}^{-1}.\nonumber 
\end{gather}
In addition, the warm plasma flow velocity in the mirror is assumed
to be
\[
u_{\mathrm{m}}=\sqrt{\frac{2}{\pi}}v_{T\mathrm{i}}=\sqrt{\frac{2T}{\pi m_{{\rm i}}}},
\]
which correspond to the gas-dynamic loss in the case of short mirrors
and a filled loss cone \citep{Ivanov2017}. In what follows, we also
consider $r\simeq a$, assuming the leading order of the thin transition
layer approximation $\lambda_{\mathrm{w}}\ll a$. 

Let us show that in this case there exists an equilibrium of the warm
plasma such that the temperature can be considered slowly varying
on the scale of the warm plasma inhomogeneity:

\begin{equation}
\frac{n_{\mathrm{i}}}{T}\frac{dT}{dn_{\mathrm{i}}}\simeq\frac{p}{T}\frac{dT}{dp}=\gamma_{T}\ll1.\label{eq:T_const}
\end{equation}
Subtracting the equation (\ref{eq:equilibrium_1_CBM}) multiplied
by $\alpha_{\mathrm{E}}T$ from the equation (\ref{eq:equilibrium_2_CBM})
yields:
\[
\alpha_{\mathrm{E}}Td\left(\Lambda_{\mathrm{i}\perp}\frac{dp}{d\psi}\right)-d\left(\frac{5\left(1+Z\right)}{2}T\Lambda_{\mathrm{i}\perp}\frac{dp}{d\psi}\right)-d\left(\Lambda_{\mathrm{E}\perp}\frac{dT}{d\psi}\right)=0.
\]
Taking into account the condition (\ref{eq:T_const}), the left-hand
side of the resulting expression is approximately reduced to an exact
differential:

\[
d\left(\left(\alpha_{\mathrm{E}}-\frac{5\left(1+Z\right)}{2}\right)T\Lambda_{\mathrm{i}\perp}\frac{dp}{d\psi}\right)-d\left(\Lambda_{\mathrm{E}\perp}\frac{dT}{d\psi}\right)\simeq0,
\]
which further allows the equation to be explicitly integrated. The
constant of integration should be set equal to zero, since at the
boundary of the warm plasma, where $n_{\mathrm{i}}\rightarrow0$,
the transverse fluxes of ions and energy should vanish. Finally, we
find the slope factor:
\begin{equation}
\gamma_{T}=\frac{p}{T}\frac{dT}{dp}\simeq\left(\alpha_{\mathrm{E}}-\frac{5\left(1+Z\right)}{2}\right)\frac{\Lambda_{\mathrm{i}\perp}}{\Lambda_{\mathrm{E}\perp}}p\simeq\frac{1}{60}\left(\alpha_{\mathrm{E}}-\frac{5\left(1+Z\right)}{2}\right)\frac{1+Z}{\left(\mu_{\mathrm{i}}^{\nicefrac{1}{2}}+0.077Z\right)Z}.\label{eq:dT}
\end{equation}
For hydrogen plasma $Z=1$, $\mu_{\mathrm{i}}=1$ and $\alpha_{\mathrm{E}}=8$,
which is typical for GDT \citep{Ivanov2017,Skovorodin2019,Soldatkina2020},
the slope factor proves to be quite small: $\gamma_{T}\sim10^{-1}\ll1$. 

When the condition (\ref{eq:T_const}) is met, the equilibrium equations
(\ref{eq:equilibrium_1_CBM}) and (\ref{eq:equilibrium_2_CBM}) are
equivalent and reduced to:
\begin{equation}
\frac{d\mathcal{F}}{d\chi}+2\beta_{\mathrm{w}}\simeq0,\quad\mathcal{F}=-\frac{\beta_{\mathrm{w}}}{R}\frac{d\beta_{\mathrm{w}}}{d\chi},\label{eq:equilibrium_CBM_TTL}
\end{equation}
where we use the dimensionless magnetic flux $\chi=\psi/\psi_{\mathrm{GD}}$
normalized to $\psi_{\mathrm{GD}}=2\pi a\lambda_{\mathrm{GD}}B_{\mathrm{v}}$,
and $\lambda_{\mathrm{GD}}$ is defined by (\ref{eq:lambda_GD}).
We also introduce the quantity $\mathcal{F}$, which has the meaning
of the warm plasma transverse flow. The magnetic field distribution
is given by (\ref{eq:R_phi_0}) taking into account the corresponding
renormalization of the magnetic flux: $\phi=\chi/\chi_{\mathrm{h}}$,
where $\chi_{\mathrm{h}}=\psi_{\mathrm{h}}/\psi_{\mathrm{GD}}=\left(1-k^{-1}\right)\rho_{\mathrm{NB}}/\lambda_{\mathrm{GD}}$
corresponds to the boundary of the hot ions $\phi=1$. When solving
the equation (\ref{eq:equilibrium_CBM_TTL}), according to the approximation
(\ref{eq:T_const}), the temperature should be considered constant
and equal to $T_{0}$. Further, a weak temperature dependence can
be found from the equality (\ref{eq:dT}):
\begin{equation}
\frac{T}{T_{0}}=\left(\frac{p}{p_{0}}\right)^{\gamma_{T}}.\label{eq:T}
\end{equation}

Internal boundary conditions (\ref{eq:flow_equilibrium_internal})
and (\ref{eq:thermal_equilibrium_internal}) also merge into one:
\begin{gather}
\left.\mathcal{F}\right|_{\chi\rightarrow0}\simeq\mathcal{S},\label{eq:equilibrium_CBM_TTL_0}\\
\mathcal{S}=\beta_{\mathrm{w}0}\frac{3}{2}\sqrt{\frac{\pi}{2}}\frac{\rho_{\mathrm{i}0}}{\lambda_{\mathrm{GD}}}\left(\frac{W_{\mathrm{i}0}}{2\Phi_{\mathrm{i}\parallel0}}-1\right),\label{eq:S}
\end{gather}
with an additional condition: 
\begin{equation}
\alpha_{\mathrm{E}}T_{0}\left(W_{\mathrm{i}0}-2\Phi_{\mathrm{i}\parallel0}\right)=Q_{\mathrm{h}0}-2\alpha_{\mathrm{E}0}T_{0}\Phi_{\mathrm{i}\parallel0},\label{eq:thermal_equilibrium_0}
\end{equation}
which defines the relationship between the temperature and the pressure
of the warm plasma in the bubble core. For simplicity, we further
assume $\alpha_{\mathrm{E}0}=\alpha_{\mathrm{E}}$, then the temperature
of the warm plasma $T_{0}$ can be explicitly found from the equality
(\ref{eq:thermal_equilibrium_0}). It proves to be independent of
the pressure and is determined by the ratio of the sources: 
\begin{equation}
T_{0}\simeq\frac{Q_{\mathrm{h}0}}{\alpha_{\mathrm{E}}W_{\mathrm{i}0}}.\label{eq:T_0}
\end{equation}

External boundary conditions (\ref{eq:equilibrium_external}) are
reduced to the following:
\begin{equation}
\left.\beta_{\mathrm{w}}\right|_{\chi=\chi_{\mathrm{w}}}=0,\quad\Leftrightarrow\quad\left.\mathcal{F}\right|_{\chi=\chi_{\mathrm{w}}}=0,\label{eq:equilibrium_CBM_TTL_external}
\end{equation}
where $\chi=\chi_{\mathrm{w}}$ corresponds to the outer boundary
of the warm plasma $r=a+\lambda_{\mathrm{w}}$. The second condition
is essentially the vanishing of the transverse flow at the boundary,
which is consistent with the pressure gradient and the current density
being finite. It is also worth noting that the position of the boundary
$\chi_{\mathrm{w}}$ is not specified and should be found self-consistently
from the solution of the equilibrium equation (\ref{eq:equilibrium_CBM_TTL}).
This means that the two conditions (\ref{eq:equilibrium_CBM_TTL_external})
are related by $\chi_{\mathrm{w}}$ and are formally reducible to
one. In other words, one of the expressions (\ref{eq:equilibrium_CBM_TTL_external})
can be considered as a boundary condition, and the other as a definition
of the boundary position $\chi=\chi_{\mathrm{w}}$.

\subsection{Numerical solution\label{subsec:numerical_solution}}

The boundary value problem (\ref{eq:equilibrium_CBM_TTL}), (\ref{eq:equilibrium_CBM_TTL_0})
and (\ref{eq:thermal_equilibrium_0}) described above turns out to
be rather complex, and to find its exact solution we use numerical
methods. 

As mentioned above, provided that the temperature is given by the
expression (\ref{eq:T_0}), the equation (\ref{eq:a}) relates the
radius of the bubble core and the pressure of warm plasma in the core:
$k=k\left(\beta_{\mathrm{w}0}\right)$. This means that $\beta_{\mathrm{w}0}$
remains the only free (i.e. unknown yet) parameter that determines
the equilibrium. In that regard, the boundary value problem before
us is convenient to reformulate as the following Cauchy problem. The
equilibrium equations (\ref{eq:equilibrium_CBM_TTL}) can be considered
as the system of ordinary differential equations for the functions
$\beta_{\mathrm{w}}=\beta_{\mathrm{w}}\left(\chi\right)$ and $\mathcal{F}=\mathcal{F}\left(\chi\right)$
with the initial conditions (\ref{eq:equilibrium_CBM_TTL_0}) and
$\left.\beta_{\mathrm{w}}\right|_{\chi=0}=\beta_{\mathrm{w}0}$. At
the same time, the quantity $\beta_{\mathrm{w}0}$ should be treated
as a variable parameter, which is found from the external boundary
condition (\ref{eq:equilibrium_CBM_TTL_external}). 

To solve the formulated problem, we apply an approach similar to the
shooting method. For a given parameter $\beta_{\mathrm{w}0}$, the
equations (\ref{eq:equilibrium_CBM_TTL_0}) are integrated by means
of the Runge-Kutta methods, starting from the corresponding initial
conditions at $\chi=0$. Computing continues until either $\beta_{\mathrm{w}}=\beta_{\mathrm{w}}\left(\chi;\beta_{\mathrm{w}0}\right)$
or $\mathcal{F}=\mathcal{F}\left(\chi;\beta_{\mathrm{w}0}\right)$
hits zero at some point $\chi=\chi_{\mathrm{w}}\left(\beta_{\mathrm{w}0}\right)$\footnote{Here, we explicitly highlight the parametric dependence of the solutions
$\beta_{\mathrm{w}}$ and $\mathcal{F}$, as well as the warm plasma
boundary position $\chi_{\mathrm{w}}$, on the parameter $\beta_{\mathrm{w}0}$. }. Thus, by varying the parameter $\beta_{\mathrm{w}0}$, we can formally
obtain the functional dependence $\chi_{\mathrm{w}}=\chi_{\mathrm{w}}\left(\beta_{\mathrm{w}0}\right)$.
On the other hand, for a given relation $\chi_{\mathrm{w}}=\chi_{\mathrm{w}}\left(\beta_{\mathrm{w}0}\right)$,
the conditions (\ref{eq:equilibrium_CBM_TTL_external}) are actually
reduced to an equation for $\beta_{\mathrm{w}0}$ on the finite interval
$0\leq\beta_{\mathrm{w}0}\leq1$. The root of this equation, which
can be found using standard numerical methods, is a true value of
the parameter $\beta_{\mathrm{w}0}$ corresponding to the real equilibrium
profile $\beta_{\mathrm{w}}=\beta_{\mathrm{w}}\left(\chi\right)$. 

\begin{figure}
\begin{centering}
\includegraphics{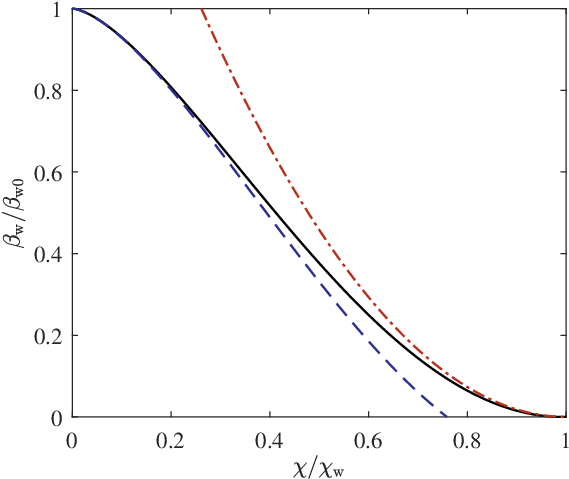}
\par\end{centering}
\caption{Profile of the warm plasma pressure depending on the magnetic flux:
the numerical solution (black solid curve), left (blue dashed curve)
and right (red dash-dotted curve) asymptotics given by the expressions
(\ref{eq:beta_left}) and (\ref{eq:beta_right}), respectively. Simulation
corresponds to the parameters of the GDMT10 regime (see Table \ref{tab:parameters}).
\label{fig:beta_chi}}
\end{figure}

Figure \ref{fig:beta_chi} shows the numerical solution of the warm
plasma equilibrium equation, found using the procedure described above.
There are also plotted the corresponding analytical asymptotics in
the vicinity of the bubble core and the outer boundary of the warm
plasma, respectively:
\begin{multline}
\beta_{\mathrm{w}}\sim\beta_{\mathrm{w}0}-\frac{4\mathcal{S}\mathfrak{C}_{\mathrm{h}}}{3\beta_{\mathrm{w}0}}\chi^{\nicefrac{3}{2}}-\frac{\mathcal{S}^{2}}{6\beta_{\mathrm{w}0}^{2}}\chi^{2}+\\
+\frac{\mathfrak{C}_{\mathrm{h}}}{5}\left\{ 8-\frac{\mathcal{S}^{3}}{36\mathfrak{C}_{\mathrm{h}}^{2}\beta_{\mathrm{w}0}^{3}}+\frac{\mathcal{S}}{\beta_{\mathrm{w}0}\chi_{\mathrm{h}}}\left[1+\left(k-1\right)\left(\ln\left(\frac{\chi}{\chi_{\mathrm{h}}}\right)-\frac{9}{10}\right)\right]\right\} \chi^{\nicefrac{5}{2}}+o\left(\chi^{\nicefrac{5}{2}}\right),\quad\chi\rightarrow0,\label{eq:beta_left}
\end{multline}
\begin{equation}
\beta_{\mathrm{w}}\sim\frac{1}{3}R\left(\chi_{\mathrm{w}}\right)\left(\chi_{\mathrm{w}}-\chi\right)^{2}+o\left[\left(\chi_{\mathrm{w}}-\chi\right)^{2}\right],\quad\chi\rightarrow\chi_{\mathrm{w}},\label{eq:beta_right}
\end{equation}
where
\begin{gather*}
\mathfrak{C}_{\mathrm{h}}=\sqrt{\frac{1}{\chi_{\mathrm{h}}}\frac{\beta_{\mathrm{h}0}}{k+1}}.
\end{gather*}
The second term in the expansion (\ref{eq:beta_left}) includes the
contribution from the hot ions – it corresponds to the pressure profile
of the warm plasma in the magnetic field induced by the hot ion current
only. The third term, in turn, takes into account the diamagnetic
current of the warm plasma. The non-polynomial logarithmic contribution
of the hot ion current appears in the fourth term of the expansion.

\section{Diamagnetic confinement in GDMT\label{sec:diamagnetic_confinement_in_gdmt}}

As an application of the constructed theoretical model, we further
investigate the equilibrium of the diamagnetic bubble corresponding
to the design parameters of the GDMT device \citep{Skovorodin2023}.
Hydrogen plasma is assumed: $Z=\mu_{\mathrm{i}}=1$; the axial loss
constants correspond to GDT \citep{Ivanov2017,Skovorodin2019,Soldatkina2020}:
$\alpha_{\mathrm{E}0}=\alpha_{\mathrm{E}}=8$; the Coulomb logarithm
is set equal to $\Lambda_{\mathrm{ie}}=15$. Energy, angle and impact
parameter of the injection are $\mathcal{E}_{\mathrm{NB}}=30\,\mathrm{keV}$,
$\xi_{\mathrm{NB}}=\pi/4$ and $r_{\mathrm{NB}}=5\,\mathrm{cm}$,
respectively, which corresponds to the minimum radius of the hot ions
equal to $\overline{r}_{\min}\simeq3.54\,\mathrm{cm}$. The bubble
length is estimated from the MHD equilibrium simulations of GDMT \citep{Khristo2022} and is equal to $L=500\,\mathrm{cm}$. The total warm plasma
source intensity is fixed at $W_{\mathrm{i}0}=5\times10^{21}\,\mathrm{sec}^{-1}$.
The magnetic field in the mirrors is $B_{\mathrm{m}}=200\,\mathrm{kG}$. 

\begin{table}
\begin{centering}
\begin{tabular}{llccc} 
\multirow{2}{*}{\textbf{Quantity}} & \multirow{2}{*}{\textbf{Units}} & \multicolumn{3}{c}{\textbf{Regime}}\tabularnewline
 &  & \textbf{GDMT05} & \textbf{GDMT10} & \textbf{GDMT20}\tabularnewline
$B_{\mathrm{v}}$ & $\mathrm{kG}$ & $5$ & $10$ & $20$\tabularnewline
$B_{\mathrm{m}}$ & $\mathrm{kG}$ & $200$ & $200$ & $200$\tabularnewline
$W_{\mathrm{i}0}$ & $\mathrm{sec}^{-1}$ & $5\times10^{21}$ & $5\times10^{21}$ & $5\times10^{21}$\tabularnewline
$Q_{\mathrm{h}0}$ & $\mathrm{MW}$ & $6.11$ & $9.75$ & $14.93$\tabularnewline
$\mathcal{E}_{\mathrm{NB}}$ & $\mathrm{keV}$ & $30$ & $30$ & $30$\tabularnewline
$\xi_{\mathrm{NB}}$ & $\ensuremath{^{\circ}}$ & $45$ & $45$ & $45$\tabularnewline
$r_{\mathrm{NB}}$ & $\mathrm{cm}$ & $5$ & $5$ & $5$\tabularnewline
$\overline{r}_{\min}$ & $\mathrm{cm}$ & $3.54$ & $3.54$ & $3.54$\tabularnewline
$L$ & $\mathrm{cm}$ & $500$ & $500$ & $500$\tabularnewline
$a$ & $\mathrm{cm}$ & $20$ & $20$ & $20$\tabularnewline
$\rho_{\mathrm{NB}}$ & $\mathrm{cm}$ & $5.01$ & $2.50$ & $1.25$\tabularnewline
$\lambda_{\mathrm{w}\,\mathrm{MHD}}$ $^{a)}$ & $\mathrm{cm}$ & $2.92$ & $1.29$ & $0.60$\tabularnewline
$\lambda_{\mathrm{h}}$ & $\mathrm{cm}$ & $7.02$ & $3.76$ & $1.97$\tabularnewline
$\lambda_{\mathrm{w}}$ & $\mathrm{cm}$ & $2.44$ & $1.07$ & $0.46$\tabularnewline
$\beta_{\mathrm{w}0}$ & $\%$ & $12.55$ & $3.66$ & $0.99$\tabularnewline
$n_{\mathrm{i}0}$ & $\mathrm{cm}^{-3}$ & $4.09\times10^{13}$ & $2.98\times10^{13}$ & $2.11\times10^{13}$\tabularnewline
$T_{0}$ & $\mathrm{keV}$ & $0.95$ & $1.52$ & $2.33$\tabularnewline
$\eta_{\parallel0}$ $^{b)}$ & $\%$ & $73.5$ & $85.6$ & $92.6$\tabularnewline
\end{tabular}{\par$^{a)}$Thickness of the warm plasma transition layer according to MHD equilibrium
\citep{Beklemishev2016}: $\lambda_{\mathrm{w}\,\mathrm{MHD}}=7\lambda_{\mathrm{GD}}$.
See the begging of Section \ref{sec:thin_transition_layer_limit}.\\
$^{b)}$Proportion of the non-adiabatic loss (\ref{eq:Phi_i_parallel_0})
in the total axial loss of the warm plasma: $\eta_{\parallel0}=2\Phi_{\mathrm{i}\parallel0}/W_{\mathrm{i}0}$.}
\par\end{centering}
\caption{Parameters of the numerical simulations. The thicknesses of transition
layers $\lambda_{\mathrm{h}}$ and $\lambda_{\mathrm{w}}$ are the
widths of the current profiles shown in Figure \ref{fig:current}.
The warm plasma thermodynamic parameters: relative pressure $\beta_{\mathrm{w}0}$,
density $n_{\mathrm{i}0}$ and temperature $T_{0}$ are the maxima
of the corresponding quantities shown in Figure \ref{fig:warm_plasma}.
The parameter $\eta_{\parallel0}$ is found from numerical simulations.\label{tab:parameters}}
\end{table}

We consider the conventional case of the vacuum magnetic field equal
to $B_{\mathrm{v}}=10\,\mathrm{kG}$, as well as the regimes with
halved and doubled fields: $B_{\mathrm{v}}=5\,\mathrm{kG}$ and $B_{\mathrm{v}}=20\,\mathrm{kG}$.
In all the simulations, the radius of the bubble core is fixed at
$a=20\,\mathrm{cm}$, which is the characteristic expected plasma
radius in the GDMT. Then the required total absorbed injection powers
prove to be $Q_{\mathrm{h}0}\simeq6.11\,\mathrm{MW}$, $9.75\,\mathrm{MW}$
and $14.93\,\mathrm{MW}$ for $B_{\mathrm{v}}=5$, $10$ and $20\,\mathrm{kG}$,
respectively. For convenience, the simulation parameters are listed
in Table \ref{tab:parameters}, where the considered regimes are briefly
called 'GDMT05', 'GDMT10' and 'GDMT20'. 

\begin{figure}
\begin{centering}
\includegraphics{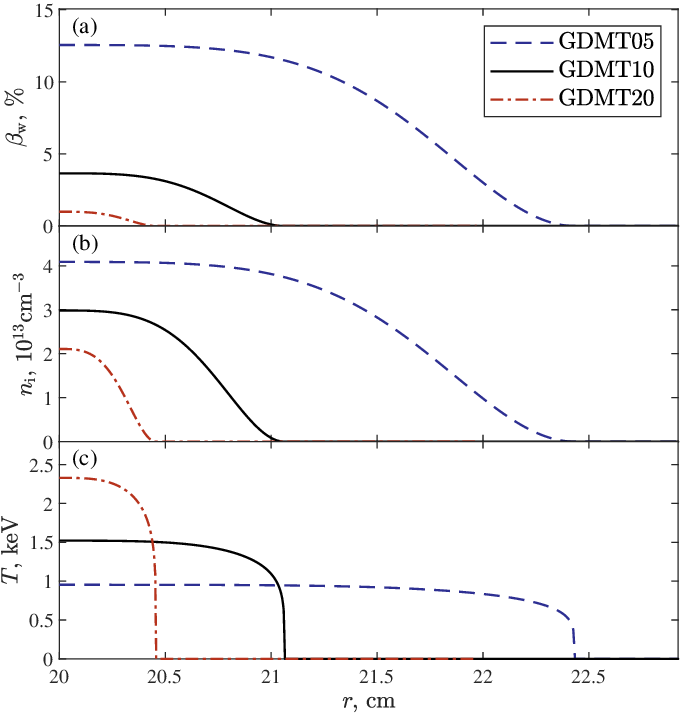}
\par\end{centering}
\caption{Radial profiles of the warm plasma (a) relative pressure, (b) density
and (c) temperature outside the bubble core $r\protect\geq a$ for
the regimes GDMT05 (blue dashed curves), GDMT10 (black solid curves)
and GDMT20 (red dash-dotted curves) specified in Table \ref{tab:parameters}.
\label{fig:warm_plasma}}
\end{figure}

\begin{figure}
\begin{centering}
\includegraphics{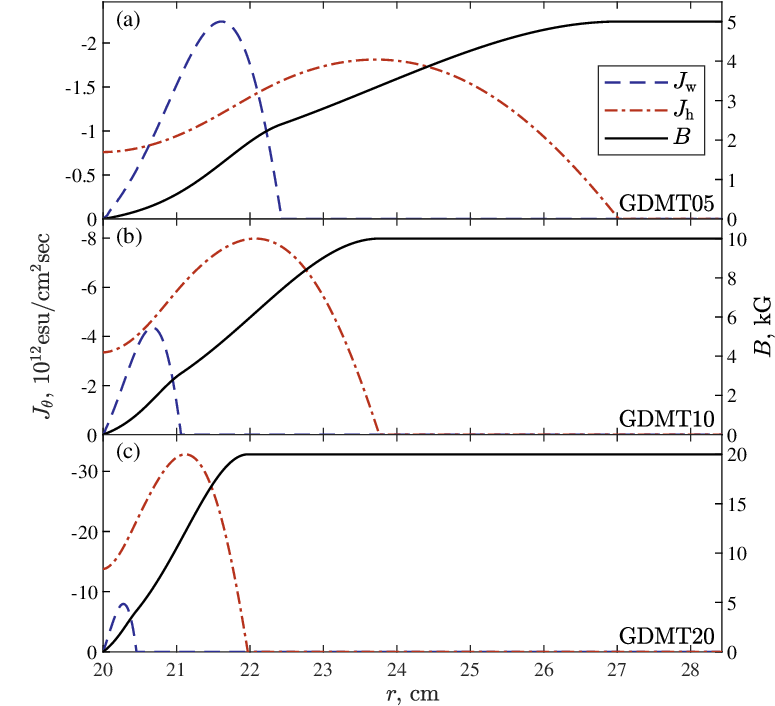}
\par\end{centering}
\caption{Equilibrium radial profiles of the magnetic field (black solid curves),
the current densities of the warm plasma (blue dashed curves) and
the hot ions (red dash-dotted curves) outside the bubble core $r\protect\geq a$.
Simulation parameters are presented in Table \ref{tab:parameters}:
(a) GDMT05, (b) GDMT10 and (c) GDMT20. \label{fig:current}}
\end{figure}

The results of the simulations are shown in Figures \ref{fig:warm_plasma}
and \ref{fig:current}. Figure \ref{fig:warm_plasma} shows the radial
profiles of the warm plasma relative pressure, density and temperature
outside\footnote{Inside the bubble core $r<a$, the warm plasma is isotropic, homogeneous
and perfectly conducting, which corresponds to the magnetic field
and the total diamagnetic plasma current being identically zero inside
the core (see Section \ref{sec:warm_plasma_equilibrium}).} the bubble core $r\geq a$. The radial profiles of the magnetic field,
along with the current densities of the warm plasma and the hot ions,
outside the core $r\geq a$ are presented in Figure \ref{fig:current}. 

\subsection*{Analysis of the solutions}
\begin{enumerate}
\item For all solutions the relative energy density of the hot ions is significantly
greater than the warm plasma relative pressure (see Table \ref{tab:parameters}
and Figure \ref{fig:warm_plasma}):
\begin{equation}
\beta_{\mathrm{w}0}\ll\beta_{\mathrm{h}0}\sim1.\label{eq:beam_regime}
\end{equation}
Thus, it turns out that the energy content of the plasma as well as
the diamagnetic current almost entirely correspond to the hot ions,
while the contribution of the warm plasma is negligible. Since the
warm plasma density in this case should be relatively small, the drag
force (\ref{eq:friction_force}) acting on the injected ions appears
to be weak as well. Therefore, this results in the energy being accumulated
by the hot component rather than being transferred to the warm plasma,
which is actually consistent with (\ref{eq:beam_regime}).
\item In Table \ref{tab:parameters}, listed are the simulated values of
the transition layer thickness for the warm plasma $\lambda_{\mathrm{w}}$
and the hot ions $\lambda_{\mathrm{h}}$, determined from the corresponding
current profiles plotted in Figure \ref{fig:current}. As expected,
the total transition layer thickness $\lambda=\max\left\{ \lambda_{\mathrm{w}},\lambda_{\mathrm{h}}\right\} $
is manly determined by the hot component, since the transition layer
for the warm plasma $\lambda_{\mathrm{w}}$ proves to be much thinner
than that for the hot ions $\lambda_{\mathrm{h}}$. The thickness
of the transition layer for the hot ions naturally turns out to be
on the order of the Larmor radius of the injected ions in the vacuum
magnetic field: $\lambda_{\mathrm{h}}\sim\rho_{\mathrm{NB}}$. At
the same time, the warm plasma transition layer thickness proves to
be surprisingly close to the estimate made for MHD equilibrium \citep{Beklemishev2016}:
$\lambda_{\mathrm{w}\,\mathrm{MHD}}=7\lambda_{\mathrm{GD}}$. This
effect is probably due to a combination of the following two factors.
On the one hand, the warm plasma transition layer should apparently
be wider in the presence of injected ions, since the characteristic
scale of the warm plasma transverse diffusion proves to be greater
in the magnetic field weakened by the diamagnetism of the hot component.
On the other hand, for a fixed warm plasma source, the lower maximum
relative pressure of the warm plasma (\ref{eq:beam_regime}) seems
to correspond to a radial pressure drop in a more narrow region. A
more clear explanation of this phenomenon and a proper quantitative
estimate of the thickness $\lambda_{\mathrm{w}}$ should be made based
on analysis the warm plasma equilibrium equations, which is planned
to be addressed in future work. 
\item The equilibria presented in Figures \ref{fig:warm_plasma} and \ref{fig:current}
are obtained within the thin transition layer limit $\lambda\ll a-\overline{r}_{\min}$
(see Section \ref{sec:thin_transition_layer_limit}). For all the
solutions found, the expansion parameter $\lambda/\left(a-\overline{r}_{\min}\right)$
proves to be less than unity. However, in the regime GDMT05 with relatively
weak vacuum magnetic field (Figure \ref{fig:current}(a)), the expansion
parameter is not too small: $\lambda/\left(a-\overline{r}_{\min}\right)\simeq7/20$.
A proper description of such equilibria should be obtained by accurate
numerical modeling that takes into account the finite thickness of
the transition layer.
\item To simplify the system of the warm plasma equilibrium equations presented
in Section \ref{subsec:warm_plasma_equilibrium}, we assume the temperature
of the warm plasma to be approximately constant: $T\simeq\mathrm{const}$.
As a result, the temperature profiles determined by the formula (\ref{eq:T})
actually prove to be relatively flat, but in addition, the corresponding
pressure and density profiles turn out to be identically shaped (see
Figure \ref{fig:warm_plasma}). It is worth noting, however, that
the approximation of a slowly varying temperature $T\simeq\mathrm{const}$
is valid in the particular case of the hydrogen plasma with the energy
loss factor $\alpha_{\mathrm{E}}$ corresponding to the GDT \citep{Ivanov2017,Skovorodin2019,Soldatkina2020},
i.e. $Z=\mu_{\mathrm{i}}=1$ and $\alpha_{\mathrm{E}}=8$. At the
same time, the nature of energy loss in the diamagnetic confinement
mode may differ significantly from the gas-dynamic regime. In particular,
the presence of the non-adiabatic loss \citep{Chernoshtanov2020,Chernoshtanov2022b}
could greatly affect both longitudinal and transverse equilibrium.
\item In Section \ref{subsec:hot_ion_equilibrium}, we assume that the injected
hot ions release the energy mainly inside the bubble core. As already
mentioned, the warm plasma temperature radial profile turns out
to be almost flat, while the density of the warm plasma $n_{\mathrm{i}}$
sharply drops just beyond the bubble core on the scale of $\lambda_{\mathrm{w}}$
(Figure \ref{fig:warm_plasma}). In the thin transition layer approximation
$\lambda_{\mathrm{w}}\leq\lambda\ll a-\overline{r}_{\min}$, this
results in the ion-electron drag $\nu_{\mathrm{ie}}\propto n_{\mathrm{i}}T^{-\nicefrac{3}{2}}$
indeed acting mainly inside the core. However, this appears to take
place only when the source of the warm plasma is entirely contained
inside the bubble core. Otherwise, the maximum density of the warm
plasma could be located outside the core, which, apparently, could
significantly increase the energy loss.
\item The simulations also yield the parameter $\eta_{\parallel0}=2\Phi_{\mathrm{i}\parallel0}/W_{\mathrm{i}0}$
(see Table \ref{tab:parameters}), which represents the total proportion
of the non-adiabatic loss (\ref{eq:Phi_i_parallel_0}) in the total
warm plasma loss. It can be seen that the warm plasma loss turns out
to be almost entirely non-adiabatic for all the considered regimes:
\begin{equation}
W_{\mathrm{i}0}\sim2\Phi_{\mathrm{i}\parallel0},\label{eq:non_adiabatic_limit}
\end{equation}
while the loss at the periphery of the bubble core $2\Phi_{\mathrm{i}\parallel\mathrm{out}}=W_{\mathrm{i}0}-2\Phi_{\mathrm{i}\parallel0}$
proves to be negligible: $2\Phi_{\mathrm{i}\parallel\mathrm{out}}\ll W_{\mathrm{i}0}$.
One can also observe that the parameter $\eta_{\parallel0}$ grows
with increasing vacuum magnetic field. The reason for this is probably
related to the high energy density of hot ions (\ref{eq:beam_regime}).
The strong diamagnetism induced by the hot ions extending beyond the
bubble core leads to a considerable weakening of the magnetic field
$B$ in the vicinity of the bubble core $r\simeq a$. Therefore, the
effective mirror ratio $\mathfrak{R}=B_{\mathrm{m}}/B$ inside the
thin transition layer $a<r<a+\lambda_{\mathrm{w}}$ is greatly increased
compared to the vacuum value $\mathcal{R}=B_{\mathrm{m}}/B_{\mathrm{v}}$.
This should eventually result in the axial loss at the periphery $\Phi_{\mathrm{i}\parallel\mathrm{out}}\sim n_{\mathrm{i}}u_{\mathrm{m}}S_{\mathrm{m}}$
being suppressed due to a decrease in the cross section of the warm
plasma flow in the mirrors $S_{\mathrm{m}}\sim a\lambda_{\mathrm{w}}/\mathfrak{R}$.
At the same time, the non-adiabatic loss $\Phi_{\mathrm{i}\parallel0}\sim n_{\mathrm{i}0}u_{\mathrm{m}0}a\rho_{\mathrm{i}0}/\mathcal{R}$,
being determined by the vacuum mirror ratio $\mathcal{R}$, is not
affected by the diamagnetism of the hot ions. 
\item The warm plasma density $n_{\mathrm{i}0}$ turns out to be considerably
lower in the regimes with a higher temperature of the warm plasma
$T_{0}$ (see Figure \ref{fig:warm_plasma} and Table \ref{tab:parameters}).
This appears to be due to the improved confinement resulting from
the suppression of the axial loss in the regimes with a lower warm
plasma temperature. When the non-adiabatic loss dominates, the warm
plasma density can be approximately estimated from (\ref{eq:non_adiabatic_limit}):
\begin{equation}
n_{\mathrm{i}0}\sim\frac{W_{\mathrm{i}0}\tau_{\mathrm{i}\parallel0}}{\pi a^{2}L}\propto\frac{W_{\mathrm{i}0}B_{\mathrm{m}}}{a}\frac{1}{T_{0}},\label{eq:n_i_0_estimate}
\end{equation}
where
\[
\tau_{\mathrm{i}\parallel0}\sim\frac{a}{\rho_{\mathrm{i}0}}\tau_{\mathrm{GD}}\propto aLB_{\mathrm{m}}\frac{1}{T_{0}}
\]
is the characteristic time of the non-adiabatic loss. It turns out
that the efficiency of the axial confinement does not depend explicitly
on the mirror ratio $\mathcal{R}$, as for a 'classical' gas-dynamic
trap $\tau_{\mathrm{GD}}\propto\mathcal{R}$, but it is enhanced with
increasing the absolute value of the mirror magnetic field $B_{\mathrm{m}}$.
In addition, the warm plasma density $n_{\mathrm{i}0}$ and the confinement
time $\tau_{\mathrm{i}\parallel0}$ indeed prove to be greater at
a lower warm plasma temperature $T_{0}$. The inverse dependence of
$n_{\mathrm{i}0}$ on $T_{0}$ is the result of the non-adiabatic
loss (\ref{eq:Phi_i_parallel_0}) being proportional to $\rho_{\mathrm{i}0}v_{T\mathrm{i}0}\propto T_{0}$.
\item It can be observed from the simulations (see Figure \ref{fig:warm_plasma}
and Table \ref{tab:parameters}) that the warm plasma temperature
$T_{0}$ proves to be higher in the regimes with a stronger vacuum
magnetic field $B_{\mathrm{v}}$. Indeed, sustaining the diamagnetic
confinement equilibrium with stronger external field $B_{\mathrm{v}}$
(at fixed mirror field $B_{\mathrm{m}}$, length $L$ and radius $a$
of the bubble core) requires greater absorbed injection power $Q_{\mathrm{h}0}$,
while the increased heating power $Q_{\mathrm{h}0}$, in turn, should
result in a corresponding rise in the equilibrium temperature of the
warm plasma $T_{0}$. This relation can be clarified by considering
the force balance in the case of beam plasma (\ref{eq:beam_regime}):
\[
\Pi_{\mathrm{h}}\sim\Pi_{\mathrm{M}},
\]
where
\[
\Pi_{\mathrm{h}}\sim-\frac{Q_{\mathrm{h}0}}{\left\langle \dot{\mathcal{E}}\right\rangle _{\Gamma}V}\sim\frac{Q_{\mathrm{h}0}}{\nu_{\mathrm{ie}0}\pi a^{2}L},\quad\Pi_{\mathrm{M}}\sim\frac{B_{\mathrm{v}}^{2}}{8\pi}
\]
are the characteristic energy densities of the hot ions and the vacuum
magnetic field, respectively. This expression roughly corresponds
to the equation (\ref{eq:a}) in the limit: $\beta_{\mathrm{h}0}\rightarrow1$,
$\beta_{\mathrm{w}0}\rightarrow0$ and $\overline{r}_{\min}/a\rightarrow0$.
Taking into account the estimate (\ref{eq:n_i_0_estimate}), we obtain
the following relation:
\begin{equation}
B_{\mathrm{v}}^{2}\sim\frac{8Q_{\mathrm{h}0}}{\nu_{\mathrm{ie}0}a^{2}L}\propto\frac{1}{aLB_{\mathrm{m}}}T_{0}^{\nicefrac{7}{2}}.\label{eq:T_0_estimate}
\end{equation}
Therefore, the temperature of warm plasma indeed proves to increase
with the vacuum magnetic field as $T_{0}\propto B_{\mathrm{v}}^{\nicefrac{4}{7}}$.
\item Having fixed the mirror field $B_{\mathrm{m}}$, the warm plasma source
$W_{\mathrm{i}0}$, the radius $a$ and the length $L$ of the bubble
core, from the considered above qualitative estimates (\ref{eq:n_i_0_estimate})
and (\ref{eq:n_i_0_estimate}), we get
\[
n_{\mathrm{i}0}\propto B_{\mathrm{v}}^{-\nicefrac{4}{7}},\quad T_{0}\propto B_{\mathrm{v}}^{\nicefrac{4}{7}}.
\]
Given the error due to the corresponding inaccuracy of the estimate
(\ref{eq:non_adiabatic_limit}), these relations agree well with the
results of the simulations shown in Figure \ref{fig:warm_plasma}
and Table \ref{tab:parameters}. 
\item In the present paper, we assume the classical model of transverse
transport for the warm plasma, which corresponds to the Spitzer conductivity
(\ref{eq:Spitzer}). At the same time, large gradients of equilibrium
parameters, such as the magnetic field and the warm plasma density,
inside the transition layer of the diamagnetic bubble could probably
lead to non-classical transport, which corresponds to a much greater
diffusion coefficient. In turn, a significant modification of the
transverse transport could apparently have a qualitative impact on
the equilibrium of the warm plasma. However, the issue concerning
non-classical diffusion in the diamagnetic confinement mode remains
poorly studied so far. 
\item The model of the warm plasma equilibrium, constructed in Section \ref{sec:warm_plasma_equilibrium}
within the framework of MHD, assumes isotropic pressure and gas-dynamic
axial loss. This is known to correspond to the collisional regime
with a filled loss cone. In the case of GDT \citep{Ivanov2017},
such a regime is realized when the gas-dynamic time $\tau_{\mathrm{GD}}\sim\mathcal{R}L/v_{T\mathrm{i}}$
significantly exceeds the kinetic time $\tau_{\mathrm{kin}}\sim\tau_{\mathrm{ii}}\ln\mathcal{R}$,
where $\tau_{\mathrm{ii}}$ is the mean free time of ion-ion Coulomb
collisions \citep{Trubnikov1965}. However, when considering a diamagnetic
trap, one should also take into account the effective increase in
the mirror ratio $\mathfrak{R}\geq\mathcal{R}$, which results from
the magnetic field weakening induced by the strong diamagnetic current
of the high-pressure plasma. In addition, the exotic conditions of
the diamagnetic confinement mode are likely to lead to anomalous collisionality
with an effective mean free time $\tau_{\mathrm{eff}}$, which typically
proves to be considerably shorter than the 'classical' time $\tau_{\mathrm{ii}}$.
Thus, the warm plasma loss cone in the transition layer of a diamagnetic
bubble can be considered filled when 
\begin{equation}
\frac{L\mathfrak{R}}{v_{T\mathrm{i}}}\gg\tau_{\mathrm{eff}}\ln\mathfrak{R}.\nonumber
\end{equation}
This condition seems to be always satisfied at least in some neighborhood
of the bubble core $r\gtrsim a$, where the magnetic field approaches
zero $B\rightarrow0$, and, accordingly, the effective mirror ratio
is greatly increased $\mathfrak{R}\rightarrow+\infty$. At the periphery
of the warm plasma $r\lesssim a+\lambda_{\mathrm{w}}$, the loss cone
may turn out to be only partially filled ($L\mathfrak{R}/v_{T\mathrm{i}}\sim\tau_{\mathrm{eff}}\ln\mathfrak{R}$)
or even empty ($L\mathfrak{R}/v_{T\mathrm{i}}\ll\tau_{\mathrm{eff}}\ln\mathfrak{R}$),
depending on the specific value of the anomalous mean free time $\tau_{\mathrm{eff}}$. 
\end{enumerate}

\section{Summary\label{sec:summary}}

In the present work, we have constructed the theoretical model of
plasma equilibrium in the diamagnetic confinement regime in an axisymmetric
mirror device with neutral beam injection. To describe the equilibrium
of the background warm plasma, we use the MHD equations of mass and
energy conservation, as well as the force balance equation. Transverse
transport is considered to be due to resistive diffusion, and the
axial loss includes both the 'classical' gas-dynamic loss and the
non-adiabatic loss \citep{Chernoshtanov2020,Chernoshtanov2022b}
specific to the diamagnetic bubble. This model does not take into
account the effects of the warm plasma rotation and inhomogeneity
of electrostatic potential. The equilibrium of the hot ions resulted
from the neutral beam injection is described by the distribution function,
which is found from the corresponding kinetic equation. It takes into
account the warm electron drag force acting on the hot ions, while
the angular scattering due to the ion-ion collisions is neglected.
The chaotic nature of the dynamics of ions in the diamagnetic bubble
is taken into consideration as well. Solving the equilibrium equations
yields the equilibrium profiles of the warm plasma density, temperature
and pressure; the hot ion current density is obtained from the corresponding
distribution function; the equilibrium magnetic field is determined
by Maxwell's equations.

Applying the constructed theoretical model, we have considered the
case of the cylindrical core of the diamagnetic bubble. In this case,
the equilibrium model is reduced to a simpler one-dimensional problem.
To further simplify the equations, we have assumed the approximation
of the thin transition layer. This allows the hot ion current density
and the magnetic field to be explicitly expressed in terms of the
magnetic flux. Finally, it has been found that the radial profile
of the warm plasma typically turns out to be almost isothermal, which
enables the system of two equilibrium equations for the warm plasma
to be approximately reduced to a single one. For the resulting equation
of the simplified equilibrium model, we have constructed a numerical
solution algorithm that includes a variation of the shooting method
in combination with the Runge-Kutta schemes. 

The equilibria of the diamagnetic bubble have been found for the parameters
of the GDMT device \citep{Skovorodin2023}. Three cases corresponding
to different vacuum magnetic fields in the central plane, $B_{\mathrm{v}}=5,\,10,\,20\,\mathrm{kG}$,
have been considered. All the other external parameters are fixed
except the absorbed injection power, which is adjusted so that the
radius of the bubble core remains the same $a=20\,\mathrm{cm}$. Equilibrium
profiles of the warm plasma density, temperature and pressure have
been found. In addition, the radial distributions of the magnetic
field and the diamagnetic current densities of the warm plasma and
the hot ions have been constructed. 

It has been found that for all the solutions obtained, the main contribution
to the plasma energy content and the diamagnetism comes from the hot
ions. This corresponds to a negligible relative pressure of the warm
plasma $\beta_{\mathrm{w}0}\ll\beta_{\mathrm{h}0}\sim1$. In addition,
the non-adiabatic loss turns out to be dominant in all considered
regimes, and its fraction in total loss is greater for stronger vacuum
magnetic field. This seems to be due to the warm plasma in the transition
layer being confined in the magnetic field weakened by the hot ion
diamagnetism, which should lead to an increased mirror ratio and improved
axial confinement. The transition layer of the diamagnetic bubble
turns out to be quite thin in the regimes with the stronger vacuum
fields, $B_{\mathrm{v}}=10,\,20\,\mathrm{kG}$. However, in the case
of the weak external field, $B_{\mathrm{v}}=5\,\mathrm{kG}$, the
thickness of the transition layer proves to be not too small, which
may correspond to a reduced accuracy of the approximate solution.
Qualitative estimates of the warm plasma density and the temperature
of the warm plasma have also been obtained. 

\subsection*{Future work}

This work should be considered as a basis for further expansion of
the theoretical model constructed here. In the near future, the equilibrium
model presented in this paper will be used for a detailed investigation
of the diamagnetic confinement regime in GDMT. In particular, it is
planned to study the dependence of the bubble equilibrium on external
conditions, such as heating power, vacuum magnetic configuration,
warm plasma source, etc. In addition, the finite absorption efficiency
of the injected neutral beam can be taken into account. The constructed
model can also be refined by eliminating the thin transition layer
approximation; this, however, would require more complex numerical
simulations. In perspective, it is also planned to include in the
model the effects of the warm plasma rotation and the inhomogeneity
of the electrostatic potential. In the long term, collisional angular
scattering of the hot ions can also be taken into account. 

\section{Acknowledgments}

The authors would like to express their gratitude to Vadim Prikhodko,
Ivan Chernoshtanov, Dmitriy Skovorodin, Elena Soldatkina, Timur Akhmetov,
Sergei Konstantinov and other colleagues from the plasma department
of the Budker INP for helpful suggestions and productive discussions.

The work was supported by the Foundation for the Advancement of Theoretical
Physics and Mathematics 'BASIS'. 

\appendix

\section{Non-adiabatic loss\label{sec:non_adiabatic_loss}}

By definition, the ion flux through the mirror throat at $z=z_{\mathrm{m}}$
is
\begin{multline*}
\Phi_{\mathrm{i}\parallel0}=\left.\left(\int dS_{\perp}\iiint\limits _{v_{z}>0}d^{3}vv_{z}f_{\mathrm{i}}\right)\right|_{z=z_{\mathrm{m}}}=\\
=\frac{n_{\mathrm{i}0}}{\sqrt{2\pi}}\left(\frac{m_{\mathrm{i}}}{T_{0}}\right)^{\nicefrac{3}{2}}\int\limits _{0}^{+\infty}rdr\int\limits _{-\infty}^{+\infty}dv_{x}\int\limits _{-\infty}^{+\infty}dv_{y}\int\limits _{0}^{+\infty}v_{z}dv_{z}\left.\left[e^{-\frac{\mathcal{E}}{T_{0}}}\Theta\left(a-\frac{\left|\mathcal{P}\right|}{\sqrt{2m_{\mathrm{i}}\mathcal{E}}}\right)\right]\right|_{z=z_{\mathrm{m}}},
\end{multline*}
where $z$ is directed along the magnetic field in the mirror throat,
and we also substituted the warm ion distribution function (\ref{eq:f_i}).
Taking into account the definitions of $\mathcal{P}$ and $\mathcal{E}$
given by (\ref{eq:P_E}), we arrive at
\[
\Phi_{\mathrm{i}\parallel0}=\frac{2n_{\mathrm{i}0}T_{0}^{-\nicefrac{3}{2}}}{\sqrt{2\pi m_{\mathrm{i}}}}\int\limits _{0}^{+\infty}dr\int\limits _{-\infty}^{+\infty}d\mathcal{P}\int\limits _{\frac{\left(\mathcal{P}-\Psi_{\mathrm{m}}\right)^{2}}{2m_{\mathrm{i}}r^{2}}}^{+\infty}d\mathcal{E}\sqrt{\frac{2\mathcal{E}}{m_{\mathrm{i}}}-\frac{\left(\mathcal{P}-\Psi_{\mathrm{m}}\right)^{2}}{m_{\mathrm{i}}^{2}r^{2}}}e^{-\frac{\mathcal{E}}{T_{0}}}\Theta\left(a-\frac{\left|\mathcal{P}\right|}{\sqrt{2m_{\mathrm{i}}\mathcal{E}}}\right),
\]
where
\[
\Psi_{\mathrm{m}}=\frac{ZeB_{\mathrm{m}}}{2c}r^{2}
\]
is the magnetic flux in the mirror. It is convenient to further use
the following dimensionless variables:
\[
\eta=\frac{\mathcal{E}}{T_{0}},\quad\zeta=\frac{\mathcal{P}}{a\sqrt{2m_{\mathrm{i}}T_{0}}},\quad\xi=\frac{r^{2}}{a^{2}}.
\]
Then the integral takes the form:
\[
I\left(\gamma\right)\overset{\mathrm{def}}{=}\frac{2\Phi_{\mathrm{i}\parallel0}}{n_{\mathrm{i}0}v_{T\mathrm{i}0}\pi a^{2}}=\left(\frac{2}{\pi}\right)^{\nicefrac{3}{2}}\int\limits _{0}^{+\infty}d\xi\int\limits _{-\infty}^{+\infty}d\zeta\int\limits _{\frac{\left(\zeta-\gamma\xi\right)^{2}}{\xi}}^{+\infty}d\eta e^{-\eta}\frac{1}{\xi}\sqrt{\eta\xi-\left(\zeta-\gamma\xi\right)^{2}}\Theta\left(1-\frac{\left|\zeta\right|}{\sqrt{\eta}}\right),
\]
where we introduced the constant 
\[
\gamma=\frac{\mathcal{R}}{2\sqrt{2}}\frac{a}{\rho_{\mathrm{i}0}},
\]
and $v_{T\mathrm{i}0}=\sqrt{T_{0}/m_{\mathrm{i}}}$ is the warm ion
thermal velocity, $\rho_{\mathrm{i}0}=v_{T\mathrm{i}0}m_{\mathrm{i}}c/eZB_{\mathrm{v}}$
is the characteristic warm ion Larmor radius in the vacuum field $B_{\mathrm{v}}$,
$\mathcal{R}=B_{\mathrm{m}}/B_{\mathrm{v}}$ is the vacuum mirror
ratio. 

Integration over $\xi$, $\zeta$, and $\eta$ is convenient to rearrange
as follows:
\[
I\left(\gamma\right)=\left(\frac{2}{\pi}\right)^{\nicefrac{3}{2}}\int\limits _{0}^{+\infty}d\eta e^{-\eta}\int\limits _{-\min\left\{ \sqrt{\eta},\frac{\eta}{4\gamma}\right\} }^{\sqrt{\eta}}d\zeta\int\limits _{x_{-}}^{x_{+}}d\xi\frac{\gamma}{\xi}\sqrt{\left(\xi_{+}-\xi\right)\left(\xi-\xi_{-}\right)},
\]
where
\[
\xi_{\pm}=\frac{2\zeta\gamma+\eta\pm\sqrt{4\zeta\gamma\eta+\eta^{2}}}{2\gamma^{2}},
\]
and $\xi_{+}>\xi_{-}>0$. Integrating over $\xi$ yields:
\[
\int\limits _{\xi_{-}}^{\xi_{+}}d\xi\frac{\gamma}{\xi}\sqrt{\left(\xi_{+}-\xi\right)\left(\xi-\xi_{-}\right)}=\frac{\pi\gamma}{2}\left(\xi_{+}+\xi_{-}-2\sqrt{\xi_{+}\xi_{-}}\right)=\pi\left(\zeta-\left|\zeta\right|+\frac{\eta}{2\gamma}\right).
\]
After integration over $\zeta$ we get:
\[
I\left(\gamma\right)=\sqrt{\frac{2}{\pi}}\left[\frac{1}{\gamma}\int\limits _{0}^{+\infty}\left(\eta^{\nicefrac{3}{2}}+\frac{\eta^{2}}{8\gamma}\right)e^{-\eta}d\eta-\int\limits _{16\gamma^{2}}^{+\infty}\left(2\eta-\frac{\eta^{\nicefrac{3}{2}}}{\gamma}+\frac{\eta^{2}}{8\gamma^{2}}\right)e^{-\eta}d\eta\right].
\]
In the limit $\gamma\gg1$ the second term in the square brackets
proves to be exponentially small. Then we obtain the asymptotic behavior
of the integral $I\left(\gamma\right)$:
\[
I\left(\gamma\right)\sim\frac{3\sqrt{2}}{4\gamma}\left(1+\frac{1}{3\sqrt{\pi}\gamma}\right)+\mathcal{O}\left(\gamma^{-2}e^{-16\gamma^{2}}\right)\sim\frac{3\sqrt{2}}{4\gamma}+\mathcal{O}\left(\gamma^{-2}\right).
\]
Restoring the dimensional values, we finally arrive at:
\[
\Phi_{\mathrm{i}\parallel0}\simeq\frac{3}{4}n_{\mathrm{i}0}v_{T\mathrm{i}0}\frac{2\pi a\rho_{\mathrm{i}0}}{\mathcal{R}}.
\]

% susie put cite commands here, don't bother with citet etc just yet.

\bibliographystyle{jpp}

\bibliography{biblio}

\begin{thebibliography}{61}
\expandafter\ifx\csname natexlab\endcsname\relax\def\natexlab#1{#1}\fi
\def\au#1{#1} \def\ed#1{#1} \def\yr#1{#1}\def\at#1{#1}\def\jt#1{\textit{#1}} \def\bt#1{#1}\def\bvol#1{\textbf{#1}} \def\vol#1{#1} \def\pg#1{#1} \def\publ#1{#1}\def\arxiv#1{#1}\def\org#1{#1}\def\st#1{\textit{#1}}

\bibitem[Bagryansky {\em et~al.\/}(2016)Bagryansky, Akhmetov, Chernoshtanov, Deichuli, Ivanov, Lizunov, Maximov, Mishagin, Murakhtin, Pinzhenin, Pikhodko, Sorokin \& Oreshonok]{Bagryansky2016}
{\sc \au{Bagryansky, P~A}, \au{Akhmetov, T~D}, \au{Chernoshtanov, I~S}, \au{Deichuli, P~P}, \au{Ivanov, A~A}, \au{Lizunov, A~A}, \au{Maximov, V~V}, \au{Mishagin, V~V}, \au{Murakhtin, S~V}, \au{Pinzhenin, E~I}, \au{Pikhodko, V~V}, \au{Sorokin, A~V} \& \au{Oreshonok, V~V}} \yr{2016} {Status of the experiment on magnetic field reversal at BINP}.  \bt{In {\em AIP Conference Proceedings\/}}, ,  \vol{vol. 1771},  \pg{p. 030015}.

\bibitem[Bagryansky {\em et~al.\/}(2011)Bagryansky, Anikeev, Beklemishev, Donin, Ivanov, Korzhavina, Kovalenko, Kruglyakov, Lizunov, Maximov, Murakhtin, Prikhodko, Pinzhenin, Pushkareva, Savkin \& Zaytsev]{Bagryansky2011}
{\sc \au{Bagryansky, P~A}, \au{Anikeev, A~V}, \au{Beklemishev, A~D}, \au{Donin, A~S}, \au{Ivanov, A~A}, \au{Korzhavina, M~S}, \au{Kovalenko, Yu~V}, \au{Kruglyakov, E~P}, \au{Lizunov, A~A}, \au{Maximov, V~V}, \au{Murakhtin, S~V}, \au{Prikhodko, V~V}, \au{Pinzhenin, E~I}, \au{Pushkareva, A~N}, \au{Savkin, V~Ya} \& \au{Zaytsev, K~V}} \yr{2011}  \at{{Confinement of Hot Ion Plasma with $\beta$ = 0.6 in the Gas Dynamic Trap}}.  \jt{Fusion Science and Technology}  \bvol{59}~(1T),  \pg{31--35}.

\bibitem[Bagryansky {\em et~al.\/}(2019)Bagryansky, Beklemishev \& Postupaev]{Bagryansky2019}
{\sc \au{Bagryansky, P~A}, \au{Beklemishev, A~D} \& \au{Postupaev, V~V}} \yr{2019}  \at{{Encouraging Results and New Ideas for Fusion in Linear Traps}}.  \jt{Journal of Fusion Energy}  \bvol{38}~(1),  \pg{162--181}.

\bibitem[Beklemishev {\em et~al.\/}(2013)Beklemishev, Anikeev, Astrelin, Bagryansky, Burdakov, Davydenko, Gavrilenko, Ivanov, Ivanov, Ivantsivsky, Kandaurov, Polosatkin, Postupaev, Sinitsky, Shoshin, Timofeev \& Tsidulko]{Beklemishev2013}
{\sc \au{Beklemishev, A}, \au{Anikeev, A}, \au{Astrelin, V}, \au{Bagryansky, P}, \au{Burdakov, A}, \au{Davydenko, V}, \au{Gavrilenko, D}, \au{Ivanov, A}, \au{Ivanov, I}, \au{Ivantsivsky, M}, \au{Kandaurov, I}, \au{Polosatkin, S}, \au{Postupaev, V}, \au{Sinitsky, S}, \au{Shoshin, A}, \au{Timofeev, I} \& \au{Tsidulko, Yu}} \yr{2013}  \at{{Novosibirsk Project of Gas-Dynamic Multiple-Mirror Trap}}.  \jt{Fusion Science and Technology}  \bvol{63}~(1T),  \pg{46--51}.

\bibitem[Beklemishev(2016)]{Beklemishev2016}
{\sc \au{Beklemishev, A~D}} \yr{2016}  \at{{Diamagnetic “bubble” equilibria in linear traps}}.  \jt{Physics of Plasmas}  \bvol{23}~(8),  \pg{082506}.

\bibitem[Beklemishev {\em et~al.\/}(2010)Beklemishev, Bagryansky, Chaschin \& Soldatkina]{Beklemishev2010}
{\sc \au{Beklemishev, Alexei~D}, \au{Bagryansky, Peter~A}, \au{Chaschin, Maxim~S} \& \au{Soldatkina, Elena~I}} \yr{2010}  \at{{Vortex Confinement of Plasmas in Symmetric Mirror Traps}}.  \jt{Fusion Science and Technology}  \bvol{57}~(4),  \pg{351--360}.

\bibitem[Belchenko {\em et~al.\/}(2018)Belchenko, Davydenko, Deichuli, Emelev, Ivanov, Kolmogorov, Konstantinov, Krasnov, Popov, Sanin, Sorokin, Stupishin, Shikhovtsev, Kolmogorov, Atlukhanov, Abdrashitov, Dranichnikov, Kapitonov \& Kondakov]{Belchenko2018}
{\sc \au{Belchenko, Yu~I}, \au{Davydenko, V~I}, \au{Deichuli, P~P}, \au{Emelev, I~S}, \au{Ivanov, A~A}, \au{Kolmogorov, V~V}, \au{Konstantinov, S~G}, \au{Krasnov, A~A}, \au{Popov, S~S}, \au{Sanin, A~L}, \au{Sorokin, A~V}, \au{Stupishin, N~V}, \au{Shikhovtsev, I~V}, \au{Kolmogorov, A~V}, \au{Atlukhanov, M~G}, \au{Abdrashitov, G~F}, \au{Dranichnikov, A~N}, \au{Kapitonov, V~A} \& \au{Kondakov, A~A}} \yr{2018}  \at{{Studies of ion and neutral beam physics and technology at the Budker Institute of Nuclear Physics, SB RAS}}.  \jt{Physics-Uspekhi}  \bvol{61}~(6),  \pg{531--581}.

\bibitem[Berk {\em et~al.\/}(1987)Berk, Wong \& Tsang]{Berk1987}
{\sc \au{Berk, H~L}, \au{Wong, H~V} \& \au{Tsang, K~T}} \yr{1987}  \at{{Theory of hot particle stability}}.  \jt{Physics of Fluids}  \bvol{30}~(9),  \pg{2681--2693}.

\bibitem[Bogoliubov \& Mitropolskii(1961)]{Bogoliubov1961}
{\sc \au{Bogoliubov, N~N} \& \au{Mitropolskii, Y~A}} \yr{1961} {\em {Asymptotic methods in the theory of non-linear oscillations}\/}.  \publ{Gordon and Breach}.

\bibitem[Boronina {\em et~al.\/}(2020)Boronina, Dudnikova, Efimova, Genrikh, Vshivkov \& Chernoshtanov]{Boronina2020}
{\sc \au{Boronina, M~A}, \au{Dudnikova, G~I}, \au{Efimova, A~A}, \au{Genrikh, E~A}, \au{Vshivkov, V~A} \& \au{Chernoshtanov, I~S}} \yr{2020}  \at{{Numerical study of diamagnetic regime in open magnetic trap}}.  \jt{Journal of Physics: Conference Series}  \bvol{1640}~(1),  \pg{012021}.

\bibitem[Braginskii(1965)]{Braginskii1965}
{\sc \au{Braginskii, S~I}} \yr{1965}  \at{{Transport Processes in a Plasma}}.  \bt{In {\em Reviews of Plasma Physics\/} (ed. \ed{M~A Leontovich})}, ,  \vol{vol.~1},  \pg{p. 205}.  \publ{New York: Consultants Bureau}.

\bibitem[Chernoshtanov(2020)]{Chernoshtanov2020}
{\sc \au{Chernoshtanov, Ivan}} \yr{2020} {Collisionless particle dynamic in an axi-symmetric diamagnetic trap},  \arxiv{arXiv: 2002.03535}.

\bibitem[Chernoshtanov {\em et~al.\/}(2023)Chernoshtanov, Efimova, Soloviev \& Vshivkov]{Chernoshtanov2023}
{\sc \au{Chernoshtanov, I}, \au{Efimova, A}, \au{Soloviev, A} \& \au{Vshivkov, V}} \yr{2023}  \at{{Fast Ion–Ion Collisions Simulation in Particle-in-Cell Method}}.  \jt{Lobachevskii Journal of Mathematics}  \bvol{44}~(1),  \pg{26--32}.

\bibitem[Chernoshtanov(2022)]{Chernoshtanov2022b}
{\sc \au{Chernoshtanov, I~S}} \yr{2022}  \at{{Collisionless Particle Dynamics in Diamagnetic Trap}}.  \jt{Plasma Physics Reports}  \bvol{48}~(2),  \pg{79--90}.

\bibitem[Chernoshtanov {\em et~al.\/}(2024)Chernoshtanov, Chernykh, Dudnikova, Boronina, Liseykina \& Vshivkov]{Chernoshtanov2024}
{\sc \au{Chernoshtanov, I~S}, \au{Chernykh, I~G}, \au{Dudnikova, G~I}, \au{Boronina, M~A}, \au{Liseykina, T~V} \& \au{Vshivkov, V~A}} \yr{2024}  \at{{Effects observed in numerical simulation of high-beta plasma with hot ions in an axisymmetric mirror machine}}.  \jt{Journal of Plasma Physics}  \bvol{90}~(2),  \pg{905900211}.

\bibitem[Dimov(2005)]{Dimov2005}
{\sc \au{Dimov, Gennadii~I}} \yr{2005}  \at{{The ambipolar trap}}.  \jt{Physics-Uspekhi}  \bvol{48}~(11),  \pg{1129--1149}.

\bibitem[Efimova {\em et~al.\/}(2020)Efimova, Dudnikova \& Vshivkov]{Efimova2020}
{\sc \au{Efimova, A.~A.}, \au{Dudnikova, G.~I.} \& \au{Vshivkov, K.~V.}} \yr{2020}  \at{{3D numerical model of diamagnetic plasma confinement in gasdynamic trap}}.  \jt{AIP Conference Proceedings}  \bvol{2312}~(1),  \pg{060001}.

\bibitem[Grad(1967)]{Grad1967}
{\sc \au{Grad, Harold}} \yr{1967} {The guiding center plasma}.  \bt{In {\em Proc. of Symposia in Applied Mathematics\/} (ed. \ed{Harold Grad})},  \st{Magneto-Fluid and Plasma Dynamics},  \vol{vol.~18},  \pg{pp. 162--248}.  \publ{Providence, Rhode Island: American Mathematical Society}.

\bibitem[Grad \& Rubin(1958)]{Grad1958}
{\sc \au{Grad, H} \& \au{Rubin, H}} \yr{1958} {Hydromagnetic equilibria and force-free fields}.  \bt{In {\em Proc. of the Second UN International Conference on the Peaceful Uses of Atomic Energy\/}},  \st{Theoretical and Experimental Aspects of Controlled Nuclear Fusion},  \vol{vol.~31},  \pg{pp. 190--197}. Geneva, Switzerland.

\bibitem[Hsiao \& Miley(1985)]{Hsiao1985}
{\sc \au{Hsiao, Ming-Yuan} \& \au{Miley, George~H}} \yr{1985}  \at{{Velocity-space particle loss in field-reversed configurations}}.  \jt{Physics of Fluids}  \bvol{28}~(5),  \pg{1440}.

\bibitem[Ioffe {\em et~al.\/}(1981)Ioffe, Kanaev, Pastukhov, Piterskii \& Yushmanov]{Ioffe1981}
{\sc \au{Ioffe, M~S}, \au{Kanaev, B~I}, \au{Pastukhov, V~P}, \au{Piterskii, V~V} \& \au{Yushmanov, E~E}} \yr{1981}  \at{{Plasma heating in a magnetic cusp confinement system without injection}}.  \jt{Soviet Journal of Experimental and Theoretical Physics Letters}  \bvol{34}~(11),  \pg{570--573}.

\bibitem[Ivanov \& Prikhodko(2017)]{Ivanov2017}
{\sc \au{Ivanov, A~A} \& \au{Prikhodko, V~V}} \yr{2017}  \at{{Gas dynamic trap: experimental results and future prospects}}.  \jt{Physics-Uspekhi}  \bvol{60}~(5),  \pg{509--533}.

\bibitem[Kaiser \& Pearlstein(1985)]{Kaiser1985}
{\sc \au{Kaiser, T~B} \& \au{Pearlstein, L~Donald}} \yr{1985}  \at{{Finite Larmor radius and wall effects on the M=1 ballooning mode at arbitrary beta in axisymmetric tandem mirrors}}.  \jt{Physics of Fluids}  \bvol{28}~(3),  \pg{1003--1005}.

\bibitem[Kaufmann {\em et~al.\/}(1970)Kaufmann, Horng \& Wolfe]{Kaufmann1970}
{\sc \au{Kaufmann, Richard~L}, \au{Horng, Jiann-Tsorng} \& \au{Wolfe, Allan}} \yr{1970}  \at{{Large-amplitude hydromagnetic waves in the inner magnetosheath}}.  \jt{Journal of Geophysical Research}  \bvol{75}~(25),  \pg{4666--4676}.

\bibitem[Khristo \& Beklemishev(2019)]{Khristo2019}
{\sc \au{Khristo, Mikhail~S} \& \au{Beklemishev, Alexey~D}} \yr{2019}  \at{{High-Pressure Limit of Equilibrium in Axisymmetric Open Traps}}.  \jt{Plasma and Fusion Research}  \bvol{14},  \pg{2403007--2403007}.

\bibitem[Khristo \& Beklemishev(2022)]{Khristo2022}
{\sc \au{Khristo, M~S} \& \au{Beklemishev, A~D}} \yr{2022}  \at{{Two-dimensional MHD equilibria of diamagnetic bubble in gas-dynamic trap}}.  \jt{Plasma Physics and Controlled Fusion}  \bvol{64}~(9),  \pg{095019}.

\bibitem[Kotelnikov(2020)]{Kotelnikov2020}
{\sc \au{Kotelnikov, Igor}} \yr{2020}  \at{{On the structure of the boundary layer in a Beklemishev diamagnetic bubble}}.  \jt{Plasma Physics and Controlled Fusion}  \bvol{62}~(7),  \pg{075002}.

\bibitem[Kotelnikov {\em et~al.\/}(2022)Kotelnikov, Zeng, Prikhodko, Yakovlev, Zhang, Chen \& Yu]{Kotelnikov2022}
{\sc \au{Kotelnikov, Igor}, \au{Zeng, Qiusun}, \au{Prikhodko, Vadim}, \au{Yakovlev, Dmitri}, \au{Zhang, Keqing}, \au{Chen, Zhibin} \& \au{Yu, Jie}} \yr{2022}  \at{{Wall stabilization of the rigid ballooning m = 1 mode in a long-thin mirror trap}}.  \jt{Nuclear Fusion}  \bvol{62}~(9),  \pg{096025}.

\bibitem[Kotelnikov(2011)]{Kotelnikov2011}
{\sc \au{Kotelnikov, I~A}} \yr{2011}  \at{{Equilibrium of a High-$\beta$ Plasma with Sloshing Ions above the Mirror Instability Threshold}}.  \jt{Fusion Science and Technology}  \bvol{59}~(1T),  \pg{47--50}.

\bibitem[Kotelnikov {\em et~al.\/}(2010)Kotelnikov, Bagryansky \& Prikhodko]{Kotelnikov2010}
{\sc \au{Kotelnikov, I~A}, \au{Bagryansky, P~A} \& \au{Prikhodko, V~V}} \yr{2010}  \at{{Formation of a magnetic hole above the mirror-instability threshold in a plasma with sloshing ions}}.  \jt{Physical Review E}  \bvol{81}~(6),  \pg{067402}.

\bibitem[Kurshakov \& Timofeev(2023)]{Kurshakov2023}
{\sc \au{Kurshakov, V~A} \& \au{Timofeev, I~V}} \yr{2023}  \at{{The role of electron current in high- $\beta$ plasma equilibria}}.  \jt{Physics of Plasmas}  \bvol{30}~(9).

\bibitem[Kuznetsov {\em et~al.\/}(2015)Kuznetsov, Passot, Ruban \& Sulem]{Kuznetsov2015}
{\sc \au{Kuznetsov, E~A}, \au{Passot, T}, \au{Ruban, V~P} \& \au{Sulem, P~L}} \yr{2015}  \at{{Variational approach for static mirror structures}}.  \jt{Physics of Plasmas}  \bvol{22}~(4),  \pg{042114}.

\bibitem[Landsman {\em et~al.\/}(2004)Landsman, Cohen \& Glasser]{Landsman2004}
{\sc \au{Landsman, A~S}, \au{Cohen, S~A} \& \au{Glasser, A~H}} \yr{2004}  \at{{Regular and stochastic orbits of ions in a highly prolate field-reversed configuration}}.  \jt{Physics of Plasmas}  \bvol{11}~(3),  \pg{947--957}.

\bibitem[Lansky(1993)]{Lansky1993}
{\sc \au{Lansky, I~M}} \yr{1993}  \bt{{On the paraxial equilibrium of the finite $\beta$ plasma in open magnetic configuration}}. {\em Tech. Rep.\/} 93-96.  \org{Budker INP}.

\bibitem[Larrabee {\em et~al.\/}(1979)Larrabee, Lovelace \& Fleischmann]{Larrabee1979}
{\sc \au{Larrabee, D~A}, \au{Lovelace, R~V} \& \au{Fleischmann, H~H}} \yr{1979}  \at{{Truncated exponential-rigid-rotor model for strong electron and ion rings}}.  \jt{Nuclear Fusion}  \bvol{19}~(4),  \pg{499--503}.

\bibitem[Lichtenberg \& Lieberman(1992)]{Lichtenberg1992}
{\sc \au{Lichtenberg, A~J} \& \au{Lieberman, M~A}} \yr{1992} {\em {Regular and Chaotic Dynamics}\/},  \st{Applied Mathematical Sciences},  \vol{vol.~38}.  \publ{New York, NY: Springer New York}.

\bibitem[Lotov(1996)]{Lotov1996}
{\sc \au{Lotov, K~V}} \yr{1996}  \at{{Spontaneous formation of zero magnetic field region near the axis of a high‐$\beta$ mirror device}}.  \jt{Physics of Plasmas}  \bvol{3}~(4),  \pg{1472--1473}.

\bibitem[Lovelace {\em et~al.\/}(1978)Lovelace, Larrabee \& Fleischmann]{Lovelace1978}
{\sc \au{Lovelace, R~V}, \au{Larrabee, D~A} \& \au{Fleischmann, H~H}} \yr{1978}  \at{{A re-analysis of exponential rigid-rotor astron equilibria}}.  \jt{Physics of Fluids}  \bvol{21}~(5),  \pg{863}.

\bibitem[Morozov \& Solov'ev(1966)]{Morozov1966}
{\sc \au{Morozov, A~I} \& \au{Solov'ev, L~S}} \yr{1966}  \at{{Motion of Charged Particles in Electromagnetic Fields}}.  \bt{In {\em Reviews of Plasma Physics\/} (ed. \ed{M~A Leontovich})}, ,  \vol{vol.~2},  \pg{pp. 201--297}.  \publ{New York: Consultants Bureau}.

\bibitem[Newcomb(1981)]{Newcomb1981}
{\sc \au{Newcomb, William~A}} \yr{1981}  \at{{Equilibrium and stability of collisionless systems in the paraxial limit}}.  \jt{Journal of Plasma Physics}  \bvol{26}~(3),  \pg{529--584}.

\bibitem[Pastukhov(1978)]{Pastukhov1978}
{\sc \au{Pastukhov, V~P}} \yr{1978}  \at{{Classical transport in electrostatically plugged magnetic confinement systems}}.  \jt{Soviet Journal of Plasma Physics}  \bvol{4}~(3),  \pg{311--316}.

\bibitem[Pastukhov(1980)]{Pastukhov1980}
{\sc \au{Pastukhov, V~P}} \yr{1980}  \at{{Anomalous electron transport in the transition layer of an electrostatically plugged magnetic mirror}}.  \jt{Soviet Journal of Plasma Physics}  \bvol{6}~(5),  \pg{549--554}.

\bibitem[Pastukhov(2021)]{Pastukhov2021}
{\sc \au{Pastukhov, V~P}} \yr{2021}  \at{{Turbulent Relaxation and Anomalous Plasma Transport}}.  \jt{Plasma Physics Reports}  \bvol{47}~(9),  \pg{892--906}.

\bibitem[Qerushi \& Rostoker(2002{\natexlab{{\em a\/}}})]{Qerushi2002a}
{\sc \au{Qerushi, Artan} \& \au{Rostoker, Norman}} \yr{2002{\natexlab{{\em a\/}}}}  \at{{Equilibrium of field reversed configurations with rotation. II. One space dimension and many ion species}}.  \jt{Physics of Plasmas}  \bvol{9}~(7),  \pg{3068--3074}.

\bibitem[Qerushi \& Rostoker(2002{\natexlab{{\em b\/}}})]{Qerushi2002}
{\sc \au{Qerushi, Artan} \& \au{Rostoker, Norman}} \yr{2002{\natexlab{{\em b\/}}}}  \at{{Equilibrium of field reversed configurations with rotation. III. Two space dimensions and one type of ion}}.  \jt{Physics of Plasmas}  \bvol{9}~(12),  \pg{5001--5017}.

\bibitem[Qerushi \& Rostoker(2003)]{Qerushi2003}
{\sc \au{Qerushi, Artan} \& \au{Rostoker, Norman}} \yr{2003}  \at{{Equilibrium of field reversed configurations with rotation. IV. Two space dimensions and many ion species}}.  \jt{Physics of Plasmas}  \bvol{10}~(3),  \pg{737--752}.

\bibitem[Rostoker \& Qerushi(2002)]{Rostoker2002}
{\sc \au{Rostoker, Norman} \& \au{Qerushi, Artan}} \yr{2002}  \at{{Equilibrium of field reversed configurations with rotation. I. One space dimension and one type of ion}}.  \jt{Physics of Plasmas}  \bvol{9}~(7),  \pg{3057--3067}.

\bibitem[Ryutov(2005)]{Ryutov2005}
{\sc \au{Ryutov, D~D}} \yr{2005}  \at{{Axial Electron Heat Loss from Mirror Devices Revisited}}.  \jt{Fusion Science and Technology}  \bvol{47}~(1T),  \pg{148--154}.

\bibitem[Sagdeev {\em et~al.\/}(1988)Sagdeev, Usikov \& Zaslavsky]{Sagdeev1988}
{\sc \au{Sagdeev, R~Z}, \au{Usikov, D~A} \& \au{Zaslavsky, G~M}} \yr{1988} {\em {Nonlinear Physics: From the Pendulum to Turbulence and Chaos}\/}. {\em Contemporary concepts in physics\/} .  \publ{Harwood Academic Publishers}.

\bibitem[Shafranov(1958)]{Shafranov1958}
{\sc \au{Shafranov, V~D}} \yr{1958}  \at{{On Magnetohydrodynamical Equilibrium Configurations}}.  \jt{Journal of Experimental and Theoretical Physics}  \bvol{6}~(3),  \pg{545--554}.

\bibitem[Skovorodin(2019)]{Skovorodin2019}
{\sc \au{Skovorodin, D~I}} \yr{2019}  \at{{Suppression of secondary emission of electrons from end plate in expander of open trap}}.  \jt{Physics of Plasmas}  \bvol{26}~(1),  \pg{012503}.

\bibitem[Skovorodin {\em et~al.\/}(2023)Skovorodin, Chernoshtanov, Amirov, Astrelin, Bagryanskii, Beklemishev, Burdakov, Gorbovskii, Kotel'nikov, Magommedov, Polosatkin, Postupaev, Prikhod'ko, Savkin, Soldatkina, Solomakhin, Sorokin, Sudnikov, Khristo, Shiyankov, Yakovlev \& Shcherbakov]{Skovorodin2023}
{\sc \au{Skovorodin, D~I}, \au{Chernoshtanov, I~S}, \au{Amirov, V~Kh}, \au{Astrelin, V~T}, \au{Bagryanskii, P~A}, \au{Beklemishev, A~D}, \au{Burdakov, A~V}, \au{Gorbovskii, A~I}, \au{Kotel'nikov, I~A}, \au{Magommedov, E~M}, \au{Polosatkin, S~V}, \au{Postupaev, V~V}, \au{Prikhod'ko, V~V}, \au{Savkin, V~Ya}, \au{Soldatkina, E~I}, \au{Solomakhin, A~L}, \au{Sorokin, A~V}, \au{Sudnikov, A~V}, \au{Khristo, M~S}, \au{Shiyankov, S~V}, \au{Yakovlev, D~V} \& \au{Shcherbakov, V~I}} \yr{2023}  \at{{Gas-Dynamic Multiple-Mirror Trap GDMT}}.  \jt{Plasma Physics Reports}  \bvol{49}~(9),  \pg{1039--1086}.

\bibitem[Soldatkina {\em et~al.\/}(2008)Soldatkina, Bagryansky \& Solomakhin]{Soldatkina2008}
{\sc \au{Soldatkina, E~I}, \au{Bagryansky, P~A} \& \au{Solomakhin, A~L}} \yr{2008}  \at{{Influence of the radial profile of the electric potential on the confinement of a high-$\beta$ two-component plasma in a gas-dynamic trap}}.  \jt{Plasma Physics Reports}  \bvol{34}~(4),  \pg{259--264}.

\bibitem[Soldatkina {\em et~al.\/}(2023)Soldatkina, Chernoshtanov, Iakovlev, Beklemishev \& Khristo]{Soldatkina2023}
{\sc \au{Soldatkina, Elena~Ivanovna}, \au{Chernoshtanov, Ivan~Sergeevich}, \au{Iakovlev, Dmitrii~Vadimovich}, \au{Beklemishev, Aleksei~Dmitrievich} \& \au{Khristo, Mikhail~Sergeevich}} \yr{2023} {Method for confining high-temperature plasma in open magnetic trap}.

\bibitem[Soldatkina {\em et~al.\/}(2020)Soldatkina, Maximov, Prikhodko, Savkin, Skovorodin, Yakovlev \& Bagryansky]{Soldatkina2020}
{\sc \au{Soldatkina, E~I}, \au{Maximov, V~V}, \au{Prikhodko, V~V}, \au{Savkin, V~Ya}, \au{Skovorodin, D~I}, \au{Yakovlev, D~V} \& \au{Bagryansky, P~A}} \yr{2020}  \at{{Measurements of axial energy loss from magnetic mirror trap}}.  \jt{Nuclear Fusion}  \bvol{60}~(8),  \pg{086009}.

\bibitem[Steinhauer(2011{\natexlab{{\em a\/}}})]{Steinhauer2011b}
{\sc \au{Steinhauer, Loren~C}} \yr{2011{\natexlab{{\em a\/}}}}  \at{{Hybrid equilibria of field-reversed configurations}}.  \jt{Physics of Plasmas}  \bvol{18}~(11),  \pg{112509}.

\bibitem[Steinhauer(2011{\natexlab{{\em b\/}}})]{Steinhauer2011a}
{\sc \au{Steinhauer, Loren~C}} \yr{2011{\natexlab{{\em b\/}}}}  \at{{Review of field-reversed configurations}}.  \jt{Physics of Plasmas}  \bvol{18}~(7),  \pg{070501}.

\bibitem[Trubnikov(1965)]{Trubnikov1965}
{\sc \au{Trubnikov, B~A}} \yr{1965}  \at{{Particle Interactions in a Fully Ionized Plasma}}.  \bt{In {\em Reviews of Plasma Physics\/} (ed. \ed{M~A Leontovich})}, ,  \vol{vol.~1},  \pg{p. 105}.  \publ{New York: Consultants Bureau}.

\bibitem[Tsurutani {\em et~al.\/}(2011)Tsurutani, Lakhina, Verkhoglyadova, Echer, Guarnieri, Narita \& Constantinescu]{Tsurutani2011}
{\sc \au{Tsurutani, Bruce~T}, \au{Lakhina, Gurbax~S}, \au{Verkhoglyadova, Olga~P}, \au{Echer, Ezequiel}, \au{Guarnieri, Fernando~L}, \au{Narita, Yasuhito} \& \au{Constantinescu, Dragos~O}} \yr{2011}  \at{{Magnetosheath and heliosheath mirror mode structures, interplanetary magnetic decreases, and linear magnetic decreases: Differences and distinguishing features}}.  \jt{Journal of Geophysical Research: Space Physics}  \bvol{116}~(A2),  \pg{A02103}.

\bibitem[Turner {\em et~al.\/}(1977)Turner, Burlaga, Ness \& Lemaire]{Turner1977}
{\sc \au{Turner, J~M}, \au{Burlaga, L~F}, \au{Ness, N~F} \& \au{Lemaire, J~F}} \yr{1977}  \at{{Magnetic holes in the solar wind}}.  \jt{Journal of Geophysical Research}  \bvol{82}~(13),  \pg{1921--1924}.

\bibitem[Zhitlukhin {\em et~al.\/}(1984)Zhitlukhin, Safronov, Sidnev \& Skvortsov]{Zhitlukhin1984}
{\sc \au{Zhitlukhin, A~M}, \au{Safronov, V~M}, \au{Sidnev, V~V} \& \au{Skvortsov, Yu~V}} \yr{1984}  \at{{Confinement of a hot plasma with $\beta$$\sim$1 in an open confinement system}}.  \jt{JETP Lett}  \bvol{39}~(6),  \pg{293--296}.

\end{thebibliography}

\end{document}